\documentclass[prd,twocolumn,amsmath,amssymb]{revtex4}

\usepackage{amsmath,amsfonts,amssymb}
\usepackage{graphicx}
\usepackage[english]{babel} 
\usepackage{color}
\usepackage{booktabs}
\usepackage{feynmp}

\usepackage{graphics}
\usepackage{rotate}
\usepackage{fancyhdr} 

\usepackage{appendix}
\usepackage{hyperref}

\usepackage{txfonts}

\newcommand{\nn}{\nonumber}
\newcommand{\be}{\begin{equation}}
\newcommand{\ee}{\end{equation}}
\def\bea{\begin{eqnarray}}
\def\eea{\end{eqnarray}}
\newcommand{\bga}{\begin{gathered}}
\newcommand{\ega}{\end{gathered}}
\newcommand{\beqa}{\begin{eqnarray}}
\newcommand{\eeqa}{\end{eqnarray}}

\newcommand{\degree}{{}^\circ}

\newcommand{\refeq}[1]{Eq.~(\ref{eq:#1})}          
\newcommand{\refeqs}[2]{Eqs.~(\ref{eq:#1})--(\ref{eq:#2})}          
          
\newcommand{\refsec}[1]{Sec.~\ref{sec:#1}}
\newcommand{\refapp}[1]{App.~\ref{app:#1}}

\def\comm#1#2{[#1,\,#2]}
\def\abker#1#2#3#4{\,{}_{#4}\mathcal{K}^{#3}_{#1#2}}
\def\ket#1{\left|\left. #1  \right.\right\rangle}
\def\bra#1{\left\langle\left. #1  \right.\right|}

\newcommand{\sCml}[3]{\,_{#1} B^{#2}_{#3}}
\newcommand{\sBml}[3]{\,_{#1} C^{#2}_{#3}}
\newcommand{\id}{{\rm d}}
\def\abkerd#1#2#3#4#5{\,{}^{#5}_{#4}\mathcal{K}^{#3}_{#1#2}}
\def\abkappa#1#2#3#4#5{\,{}_{#4}\kappa^{#3(#5)}_{#1#2}}
\def\abkappab#1#2#3#4#5{\,{}_{#4}\bar{\kappa}^{#3(#5)}_{#1#2}}
\newcommand{\Kern}[2]{{}_s \mathcal{K}^m_{#1\,#2}}
\newcommand{\Kernb}[2]{{}_s \bar{\mathcal{K}}^m_{#2\,#1}}
\newcommand{\Kerns}[4]{{}_{#1} \mathcal{K}^{#3}_{#4 #2}}
\newcommand{\Kernsb}[4]{{}_{#1} \bar{\mathcal{K}}^{#3}_{#4 #2}}

\newcommand{\sYlm}[3]{{}_{#1} Y_{#2 #3}}
\newcommand{\sPlm}[3]{{}_{#1} P_{#2}^{#3}}
\newcommand{\expf}[1]{{\rm e}^{#1}}
\newcommand{\Cfunc}[1]{{}_s B^m_{#1}}
\newcommand{\vect}[1]{\ensuremath{\boldsymbol{#1}}}
\newcommand{\vecth}[1]{\ensuremath{\boldsymbol{\hat{#1}}}}

\begin{document}

\title{New operator approach to the CMB aberration kernels in harmonic space}
\author{Liang Dai and Jens Chluba}
\affiliation{Department of Physics \& Astronomy, Bloomberg Center, The Johns Hopkins University, Baltimore, MD 21218, USA}

\date{\today}

\begin{abstract}
Aberration kernels describe how harmonic-space multipole coefficients of cosmic microwave background (CMB) observables transform under Lorentz boosts of the reference frame. For spin-weighted CMB observables, transforming like the CMB temperature (i.e. Doppler weight $d=1$), we show that the aberration kernels are the matrix elements of a unitary boost operator in harmonic space. Algebraic properties of the rotation and boost generators then give simple, exact recursion relations that allow us to raise or lower the multipole quantum numbers $\ell$ and $m$, and the spin weight $s$. Further recursion relations express kernels of other Doppler weights $d\neq 1$ in terms of the $d=1$ kernels. From those we show that on the full sky, to all orders in $\beta=\varv/c$, $E$- and $B$-mode polarization observables do not mix under aberration if and only if $d=1$. The new relations, fully non-linear in the boost velocity $\beta$, form the basis of a practical recursive algorithm to accurately compute any aberration kernel. In addition, we develop a second, fast algorithm in which aberration kernels are obtained using a set of ordinary differential equations. This system can also be approximately solved at small scales, providing simple asymptotic formulae for the aberration kernels. The results of this work will be useful for further studying the effect of aberration on future CMB temperature and polarization analysis, and might provide a basis for relativistic radiative transfer schemes.
\end{abstract}
\maketitle

\section{Introduction}
\label{sec:intro}
The temperature and polarization anisotropies of the cosmic microwave background (CMB) radiation provide a great deal of information about the origin and evolution of our Universe~\cite{Smoot1992, Mather1994, Hinshaw:2012aka,Ade:2013zuv}. Inflation predicts that the primordial CMB fluctuations have isotropic and gaussian statistics around an average temperature of $\bar T=2.7260 \pm 0.0013~\mathrm{K}$~\cite{Fixsen:1996nj, Fixsen2009} in the CMB rest frame. The anomalously large temperature dipole ($\ell=1$) $\Delta T=3.355 \pm 0.008~\mathrm{mK}$~\cite{dipole} towards Galactic coordinates $(l,b)=(263.99\degree\pm0.14\degree,48.26\degree\pm0.03\degree)$~\cite{Hinshaw:2008kr}, however, indicates that the solar system is moving with respect to the CMB rest frame with a speed $\beta=\varv/c=0.00123$. Therefore, due to the Lorentz boost from the CMB rest frame into our frame, the observed radiation deviates from what would be seen in the CMB rest frame.

In addition to the change of the photon energy caused by the Doppler effect (leading to the temperature dipole), due to light aberration the photon's apparent propagation direction is also modified under a Lorentz boost (and so are the polarization direction and plane). This induces coherent, (nearly) dipolar departures from statistical isotropy in both the temperature and the polarization field. Although the Doppler and aberration effects occur independently, by {\it aberration} we henceforth refer to both of them simultaneously, unless stated otherwise.

Aberration-induced off-diagonal elements in the CMB covariance matrix can serve as an independent handle to determine the observer's motion~\cite{Challinor:2002zh, Burles:2006xf,Kosowsky:2010jm,Amendola:2010ty,Aghanim:2013suk}. The motion-induced distortion of the CMB statistics should be corrected for before accurate cosmological information can be extracted from the observed temperature/polarization power spectra. Although Ref.~\cite{Challinor:2002zh} first found that the correction is $\mathcal{O}(\beta^2)\sim 10^{-6}$ for the idealistic full-sky situation, it was later on realized that in practice the bias can be $\mathcal{O}(\beta)\sim 10^{-3}$ due to asymmetric sky masks~\cite{Pereira:2010dn,Catena:2012hq,Jeong:2013sxy,Jeong:future}. Moreover, current or incoming experiments with high resolution, e.g. Planck~\cite{Ade:2013kta}, SPT~\cite{Story:2012wx}, ACT~\cite{Fowler:2010cy}, ACTpol~\cite{Niemack:2010wz} and SPTpol~\cite{Austermann:2012ga}, push the investigation of the aberration effects to even larger multipole $\ell$ (i.e., smaller scales). This will be particularly important for polarization data, which encode primordial information to larger $\ell$~\cite{Austermann:2012ga,Niemack:2010wz}. All those aspects call for modelling the aberration effects, for both temperature and polarization anisotropies, on very small angular scales and with great precision.

One could in principle undo the aberration effects by ``de-boosting'' the sky in real space~\cite{Menzies2005, Notari:2011sb,Yoho:2012am}. In reality, however, real-space methods suffer from inaccuracies due to the resolution of the pixelization scheme and imperfect knowledge of the window function, because aberration does not preserve the shape and the area of each pixel. This also causes changes to the effective beam of the instrument that have to be considered carefully. 
To avoid these problems, Ref.~\cite{Jeong:2013sxy} proposed a harmonic-space strategy in which one first boosts the full sky in harmonic space and then transforms into real space to apply the sky mask. The precision of the harmonic-space approach is then guaranteed by accurate determination of the {\it aberration kernels} --- the linear transformation from multipole coefficients in the rest frame to those in the observer's frame. 

The aberration kernels depend not only on the spherical harmonic multipole numbers $\ell,m$, but also on the spin weight $s$ ($s=0$ and $s=\pm2$, for temperature and polarization, respectively). Furthermore, they differ for different Doppler weight $d$ (which is the power of Doppler factor present in the transformation rules), depending on whether the thermodynamic temperature ($d=1$), the specific intensity ($d=3$) or the frequency-integrated intensity ($d=4$) are being boosted \cite{Challinor:2002zh, Amendola:2010ty}. Below the typical angular scale of aberration, corresponding to $\ell \gtrsim 1/\beta \simeq 800$ or $\delta \theta \simeq 4'$, analytical results up to $\mathcal{O}(\beta^2)$~\cite{Challinor:2002zh} for the kernels break down~\cite{Chluba:2011zh}, and algorithms non-perturbative in $\beta$ are needed. General integral expressions for the kernels have been known, but their highly oscillatory nature makes direct numerical integration unfeasible. The first efficient algorithm for computation of the kernel elements based on recursions was developed in Ref.~\cite{Chluba:2011zh} to push into the non-perturbative regime. Fitting formula for the kernel integrals, approximately valid at small angular scales and tested up to intermediate $\ell\lesssim 700$, were given in Ref.~\cite{Notari:2011sb} to go beyond a power expansion in $\beta$.

In this work, we take a more systematic route than previous studies. We show that the $d=1$ kernels are the matrix elements of a unitary boost operator, analogous of the Wigner $D$-functions being the matrix elements of a rotation operator in harmonic space. The unitary operator lives in the Hilbert space of all spin-weighted functions on the sky. It is the exponentiation of the boost generator (valid for infinitesimal boost), parameterized by the rapidity parameter $\eta=\tanh^{-1}\beta$ that is additive under successive boosts. Using rapidity instead of $\beta$ to describe the boost is one of the new insights into the problem that allowed us to generalize previous discussions.
The Lorentz algebra, formed by the generators of rotation and boost in harmonic space, then leads to simple linear recursions that relate kernels of different $\ell$, different $m$ and general spin weight $s$. 
In particular, these expressions are more compact than those given in Ref.~\cite{Chluba:2011zh} and do not require an order-by-order treatment.
Moreover, the $d \neq 1$ kernels can be obtained from those of $d=1$ through another set of straightforward recursions. This is particularly interesting since the $d=1$ kernels follow special symmetries that ease their computation.

Based on our novel representation of aberration kernels, we obtain two efficient and accurate algorithms to cross check against each other: (i) an elegant recursive algorithm that improves upon Ref.~\cite{Chluba:2011zh} and accounts for kernels of arbitrary $s$ and $d$ (see \refsec{recur}); (ii) a scheme in which kernels are computed using ordinary differential equations (ODEs) as flows in the rapidity $\eta$ (\refsec{ode-rep}). We explain how to implement both algorithms as powerful solutions to boosting the sky in harmonic space. The ODE approach furthermore allows us to derive simple analytic approximations valid at small-scales. The expressions are very similar to those obtained by Ref.~\cite{Notari:2011sb}, however, here we derive them from analytic considerations also improving the range of applicability and testing them to very small scales (\refsec{asymp_exp}). Our code will be available at \url{www.Chluba.de/Aberration}.

This paper is organized as follows. \refsec{aber-kernel} reviews the definitions of harmonic-space aberration kernels (for general $s$ and $d$), their integral representations, and their basic properties. In \refsec{matrix-element-rep}, we introduce operators that generate a Lorentz boost in harmonic space, and derive the matrix-element representations for the $d=1$ kernels. In \refsec{EB-mixing}, we show that aberration does not generate mixing between $E$ and $B$ modes for polarization observables with Doppler weight $d=1$. Then in \refsec{recur}, based on the matrix-element representation, we make use of operator algebra and derive recursion relations that relate $d=1$ kernels with adjacent values of $\ell$, $m$, and the spin weight $s$. Immediately following those recursion relations, a practical recursive algorithm to compute the aberration kernels needed is then presented in detail. An alternative method based on solving ODEs, also derived from the operator approach, is developed in \refsec{ode-rep}. We offer some concluding remarks in \refsec{concl}. In \refapp{integral-form}, we include a covariant derivation of the integral forms for both temperature and polarization kernels (we illustrate by the $d=1$ case, but the derivation can be easily generalized to $d \neq 1$). Some symmetry properties of the kernels are proved in \refapp{proofs}. \refapp{Yz-action} is a brief derivation of how the boost generator acts on spherical harmonic base functions. \refapp{sign-spinweight} details a key steps used to prove the conclusion of \refsec{EB-mixing}. \refapp{sP_lm} elaborates on a few numerical techniques that provide initial conditions for our recursive algorithm and quadrature.

\section{Aberration in harmonic space}
\label{sec:aber-kernel}
In this Section, we review the harmonic-space aberration kernels. Imagine an observer that moves with respect to the CMB rest frame $S$. Without loss of generality, the spatial coordinates can be oriented such that he is moving along the $z$ direction relative to $S$ with speed $\beta = \varv/c$. We call the rest frame of the observer $S'$.

In frame $S$, a CMB observable can be defined as a function of the direction $\vecth{n}$ in the sky $X=X(\vecth{n})$. It can be the blackbody temperature of the radiation $T(\vecth{n})$, which has spin weight $s=0$ under a rotation about the line of sight. For polarization the observables are the {\it temperature-weighted} Stokes parameters, $P_{\pm}=T(\vecth{n}) [Q(\vecth{n})\mp i U(\vecth{n})]/[\sqrt 2 I(\vecth{n})]$, which have spin weight $s=\pm2$, respectively. A spherical harmonic expansion can be applied to these observables,
\bea
\label{eq:harmonic-expansion}
X(\vecth{n}) = \sum_{\ell m}\, a^X_{\ell m}\,{}_{-s_X}Y_{\ell m}(\vecth{n}),
\eea
where $a^X_{\ell m}=\int  {}_{-s_X}Y^\ast_{\ell m}(\vecth{n}) X(\vecth{n})\,\id^2 \vecth{n}$. Aberration is the phenomenon in which a photon coming from direction $\vecth{n}$ in $S$ will appear to have come from a different direction $\vecth{n}'$ in $S'$, and in addition that its energy undergoes a Doppler shift. The spherical-polar coordinates in both frames, $\vecth{n}=(\theta,\phi)$ and $\vecth{n}'=(\theta', \phi')$, are related by
\bea
\label{eq:direction-shift}
\cos\theta = \frac{\cos\theta'-\beta}{1-\beta\cos\theta'},\qquad \phi=\phi'.
\eea
Therefore, the CMB observables, $X'=X'(\vecth{n}')$, as measured in $S'$, differ from those measured in $S$. Another harmonic expansion similar to \refeq{harmonic-expansion} can be conducted, and multipole coefficients $a'^X_{\ell'm'}$ measured in $S'$ are then obtained. Those are related to the the multipole coefficients $a^X_{\ell m}$ through a linear transformation,
\bea
\label{eq:aber-kernel-def}
a'^X_{\ell'm} = \sum_{\ell} \abker{\ell'}{\ell}{m}{s_X}(\beta) \,a^X_{\ell m},
\eea
where $\abker{\ell'}{\ell}{m}{s_X}(\beta)$ is called the aberration kernel for spin weight $s_X$. Note that for observer's velocity in the $z$ direction, aberration does not mix multipoles with different $m$'s. 

Based on the transformation properties of photon's energy and its polarization tensor under a Lorentz boost, explicit expressions for aberration kernels for both temperature $T$ and polarizations $P_\pm$ have been derived in the literature in the form of an angular integral involving two (spin-weighted) spherical harmonics~\cite{Challinor:2002zh}, 
\bea
\label{eq:kernel-integral}
\abker{\ell'}{\ell}{m}{s}(\beta) & = & \int \id^2 \vecth{n}' \frac{[{}_{-s}Y_{\ell'm}(\vecth{n}')]^*\,{}_{-s}Y_{\ell m}(\vecth{n})}{\gamma(1-\beta\cos\theta')}
\eea
with Lorentz factor $\gamma=1/\sqrt{1-\beta^2}$. Here one has to view $\vecth{n}$ as a function of $\vecth{n}'$ by inserting \refeq{direction-shift}. The temperature kernels ($s=0$) is easy to derive from the relation $T'(\vecth{n}')=T(\vecth{n})/[\gamma(1-\beta \cos\theta')]$. For interested readers, the derivation giving the polarization kernels ($s=\pm2$) is outlined in \refapp{integral-form-polarization}. 
Unfortunately, the kernel integral \refeq{kernel-integral} is numerically difficult to compute (especially for large $\ell,\ell'$) due to the fast oscillatory behaviors of (spin-weighted) spherical harmonics. For this reason, recursion relations have been developed as alternative method to compute the kernels accurately and efficiently~\cite{Chluba:2011zh}. Here, we generalize and improve these recursion to $s \neq 0$.

\subsection{Generalization to $F'(\vecth{n}')/{\nu'}^d\equiv F(\vecth{n})/\nu^d$}

The kernel \refeq{kernel-integral} is applicable to any spin-weighted field on the sphere transforming as $F'(\vecth{n}')\equiv F(\vecth{n})/[\gamma(1-\beta \cos\theta')]$ or equivalently $F'(\vecth{n}')/\nu'\equiv F(\vecth{n})/\nu$ under Lorentz boosts. Here $\nu$ and $\nu'$ are the photon's frequencies as measured in $S$ and $S'$ respectively. It is straightforward to generalize to fields that transform as $F'(\vecth{n}')/{\nu'}^d\equiv F(\vecth{n})/\nu^d$, with any (integer) Doppler weight $d$, once $\abker{\ell'}{\ell}{m}{s}(\beta)$ for $d=1$ is known. Defining
\bea
\label{eq:kernel-integral_d}
\abkerd{\ell'}{\ell}{m}{s}{d}(\beta) & = & \int \id^2 \vecth{n}' \frac{[{}_{-s}Y_{\ell'm}(\vecth{n}')]^*\,{}_{-s}Y_{\ell m}(\vecth{n})}{[\gamma(1-\beta\cos\theta')]^d}
\eea
and using the identity $\gamma^2 (1-\beta \cos\theta')(1+\beta\cos\theta)=1$, with the relation \refeq{sYlm_relations} we have
\begin{align}
\label{eq:kernel-integral_d_rec_p}
\abkerd{\ell'}{\ell}{m}{s}{d}
&=\int \frac{\id^2 \vecth{n}' [{}_{-s}Y_{\ell'm}(\vecth{n}')]^*\,{}_{-s}Y_{\ell m}(\vecth{n})}{[\gamma(1-\beta\cos\theta')]^{d-1} \gamma(1-\beta\cos\theta')}
\nn\\[2mm]
&= \int \frac{\id^2 \vecth{n}' [{}_{-s}Y_{\ell'm}(\vecth{n}')]^*\,{}_{-s}Y_{\ell m}(\vecth{n})\gamma(1+\beta\cos\theta)}{[\gamma(1-\beta\cos\theta')]^{d-1}}
\nn\\[2mm]
&=\gamma \abkerd{\ell'}{\ell}{m}{s}{d-1}
+\gamma \beta
\bigg[
\sBml{s}{m}{\ell+1}  \abkerd{\ell'}{\ell+1}{m}{s}{d-1} 
\bigg.
\nn\\
&\quad\qquad\left. + \frac{sm}{\ell(\ell+1)}\abkerd{\ell'}{\ell}{m}{s}{d-1}
+ \sBml{s}{m}{\ell} \abkerd{\ell'}{\ell-1}{m}{s}{d-1}
\right],
\end{align}
where $\sBml{s}{m}{\ell}=\sqrt{(\ell^2-m^2)(\ell^2-s^2)/(4\ell^2-1)}/\ell$ for $\ell>0$ and $\ell\geq |m|,|s|$, but zero otherwise. With this expression one can raise $d$ by unity, providing recursions that can be started from $\abker{\ell'}{\ell}{m}{s}=\abkerd{\ell'}{\ell}{m}{s}{1}$. 
Similarly, to lower $d$ we can use
\begin{align}
\label{eq:kernel-integral_d_rec_m}
\abkerd{\ell'}{\ell}{m}{s}{d}
&= \int \frac{\id^2 \vecth{n}' [{}_{-s}Y_{\ell'm}(\vecth{n}')]^*\,{}_{-s}Y_{\ell m}(\vecth{n}) \gamma(1-\beta\cos\theta')}{[\gamma(1-\beta\cos\theta')]^{d+1} }
\nn\\[2mm]
&=\gamma \abkerd{\ell'}{\ell}{m}{s}{d+1}
-\gamma \beta
\bigg[
\sBml{s}{m}{\ell'+1}  \abkerd{\ell'+1}{\ell}{m}{s}{d+1} 
\bigg.
\nn\\
&\quad\quad\left.+ \frac{sm}{\ell'(\ell'+1)}\abkerd{\ell'}{\ell}{m}{s}{d+1}
+ \sBml{s}{m}{\ell'} \abkerd{\ell'-1}{\ell}{m}{s}{d+1}
\right].
\end{align}
These two relations are useful because, as we show below, the kernel for the case $d=1$ has special symmetry properties that ease its calculation. The kernel for any other Doppler weight $d\neq 1$ is then readily obtained with Eq.~\eqref{eq:kernel-integral_d_rec_p} and \eqref{eq:kernel-integral_d_rec_m}.

\subsection{General properties of the kernel}
\label{sec:kernel_properties}

Using the properties of the spin-weighted spherical harmonic functions, with the definition of the kernel integrals it is straightforward to show that $\abkerd{\ell'}{\ell}{m}{s}{d}$ has the following general properties (see Appendix~\ref{app:proofs})
\begin{subequations}
\label{eq:kernel_props}
\begin{align}
\abkerd{\ell}{\ell'}{m}{s}{d}(\beta) &= (-1)^{\ell+\ell'} \abkerd{\ell'}{\ell}{m}{-s}{2-d}(\beta)
\\
\abkerd{\ell'}{\ell}{m}{s}{d}(-\beta) &= (-1)^{\ell+\ell'}\abkerd{\ell'}{\ell}{m}{-s}{d}(\beta)
\\
\abkerd{\ell'}{\ell}{-m}{-s}{d}(\beta) &= [\abkerd{\ell'}{\ell}{m}{s}{d}(\beta)]^\ast \equiv  \abkerd{\ell'}{\ell}{m}{s}{d}(\beta) 
\end{align}
\end{subequations}
These properties highlight useful symmetries of the kernels. For instance, combining the first two equations for the case $d=1$ gives
\bea
\abker{\ell}{\ell'}{m}{s}(\beta) = \abker{\ell'}{\ell}{m}{s}(-\beta).
\eea
This means that the kernels are unitary and the total (temperature or polarization) power is conserved under a Lorentz boost~\cite{Jeong:2013sxy}. The last expression furthermore emphasizes that the aberration kernel (in the special coordinate system where $\vect{\beta}$ is aligned with the $z$-axis) is a real quantity. In particular, for $s=0$ it implies that only elements for $m\geq 0$ have to be computed, immediately determining those for $m<0$. This just reflects the fact that a map of a real quantity remains real under aberration and can be used to simplify the computation. For $s \neq 0$ the situation is more involved, and properties of the kernel under parity transformations depend on $d$, as we explain in Sec.~\ref{sec:EB-mixing}.

\section{Matrix-element representation for the aberration kernels}
\label{sec:matrix-element-rep}

To generalize our discussion to arbitrary spin weight $s$, in this section we recast \refeq{kernel-integral} into a different form using Wigner $D$-functions.

\subsection{Hilbert space for functions with spin}

While scalar fields on the sky depend on the two spherical angles $\phi$ and $\theta$, fields with general spin weight depend on an additional roll angle $\psi$ (that goes from $0$ to $2\pi$), which parameterizes a rotation about the normal direction at every point on the sky. Consider the Hilbert space of all fields with general spin weight, i.e. all functions of $\phi$, $\theta$ and $\psi$. These angles can also be thought as three Euler angles describing the orientation of a rigid body.
For two functions (or two states) $f=f(\phi,\theta,\psi)$ and $g=g(\phi,\theta,\psi)$ living in this Hilbert space, we borrow the notation from quantum mechanics and define the overlap between the two as
\bea
\label{eq:inner-product}
\left\langle f | g \right\rangle \equiv \int \id^3 \rho \,[f(\phi,\theta,\psi)]^* g(\phi,\theta,\psi),
\eea
where the integral over three angles is explicitly
\bea
\int \id^3\rho \equiv \int^{2\pi}_0 \id\phi \int^{\pi}_0 \sin\theta \id\theta \int^{2\pi}_0 \id\psi. 
\eea
Similarly, a linear operator $\hat{O}$ defined in this Hilbert space has a matrix element between the two states $\bra{f} \hat{O} \ket{g}$.

A complete set of base functions for this Hilbert space are the familiar Wigner $D$-functions, which are related to the spin-weighted harmonics via~\cite{Goldberg:1967spin-s}
\bea
D^{\ell}_{sm}(\phi,\theta,\psi) & = & \sqrt{\frac{4\pi}{2\ell+1}} \,\expf{is\psi}\,{}_{-s}Y_{\ell m}(\theta,\phi).
\eea
It is more convenient to use the normalized base functions $\tilde D^{\ell}_{sm}(\phi,\theta,\psi)=\sqrt{(2\ell+1)/(8\pi^2)} D^{\ell}_{sm}(\phi,\theta,\psi)$. Sometimes it is more compact to use the common bra-ket notation $\ket{slm}$ to denote the base functions $\tilde D^{\ell}_{sm}$. 

\subsection{Aberration kernels as matrix elements}

Using base functions $\tilde D^\ell_{sm}$, \refeq{kernel-integral} can be unified as
\begin{align}
\label{eq:kernel-integral-Dlsm}
\abker{\ell'}{\ell}{m}{s}(\beta) & = \int \id^3 \rho' \,\frac{[\tilde D^{\ell'}_{sm}(\phi',\theta',\psi')]^* \tilde D^{\ell}_{sm}(\phi,\theta,\psi) }{\gamma (1-\beta\cos\theta')}.
\end{align}
Of course, since we choose to use the variables in frame $S'$, $(\phi,\theta,\psi)$ must be viewed as functions of $(\phi',\theta',\psi')$; Since a boost preserves the spin weight, $\psi=\psi'$ can be supplemented to \refeq{direction-shift}.

We are now in a position to show that the aberration kernels are equal to the matrix elements of a unitary operator that represents the Lorentz boost. Consider first an infinitesimal boost. The relative speed $\beta=\varv/c$ corresponds to the {\it rapidity} $\eta=\tanh^{-1} \beta$, and we assume that $\eta$ is sufficiently small so that it suffices to compute up to linear order in $\eta$. Now we insert \refeq{direction-shift} and $\beta=\tanh\eta$ into \refeq{kernel-integral-Dlsm}, and expand up to linear order in $\eta$, giving
\begin{align}
\label{eq:kernal-mat-infinitesimal}
\abker{\ell'}{\ell}{m}{s}(\beta) 
&= \int \id^3 \rho' \,[\tilde D^{\ell'}_{sm}]^* \tilde D^{\ell}_{sm} 
\nn\\
&\!\!\!\!\!\!\!\!\!\!\!\!
+ \eta \int \id^3 \rho' \,[\tilde D^{\ell'}_{sm}]^* \left( \cos\theta' + \sin\theta' \frac{\partial}{\partial\theta'} \right) \tilde D^{\ell}_{sm} + \cdots \nn\\
&= \bra{sl'm} \left[ 1 + i \eta \hat Y_z + \mathcal{O}(\eta^2) \right]  \ket{slm}.
\end{align}
Since we have expanded in $\eta$, the integrands are now calculated along line of sight direction $(\phi',\theta',\psi')$ in $S'$, and the integrals evaluate to a matrix element. The differential operator $\hat Y_z$ is the {\it boost generator} along the $z$ direction,
\bea
\label{eq:Yz-explicit}
\hat Y_z(\phi,\theta,\psi,\partial_\phi,\partial_\theta,\partial_\psi) = - i \left( \cos\theta + \sin\theta \partial_\theta \right).
\eea
It is independent of the azimuthal angle and the roll angle because a boost along the $z$ direction leaves those two variables unchanged.

Generalization of \refeq{kernal-mat-infinitesimal} to finite $\eta$ is straightforward. The rapidity is additive under successive boosts. A boost with finite $\eta$ can be achieved by successively applying many boosts along the same direction but each with a very small rapidity parameter. For instance, we can take $N$ successive boosts, each with rapidity $\eta/N$. The operator for the finite boost is therefore the exponentiation of the infinitesimal one $\lim_{N\rightarrow \infty} ( 1 + i\eta \hat Y_z / N )^N = e^{i\eta \hat Y_z}$. Hence, we have
\bea
\label{eq:kernel-mat-zboost}
\abker{\ell'}{\ell}{m}{s}(\beta) = \bra{s\ell'm} e^{i\eta \hat Y_z} \ket{s\ell m},
\eea
which provides an alternative way to express the kernel matrix elements, Eq.~\eqref{eq:kernel-integral}, but using the language of operators in Hilbert space.

\subsection{Arbitrary boost direction}

Frame $S'$ can in general move along other directions relative to frame $S$. To generalize the boost direction from the $z$ direction to any other direction, we re-orient the spatial coordinate system. This can be done using the three angular momentum operators $\hat L_a$ with $a=x,y,z$~\cite{Goldberg:1967spin-s},
\bea
\label{eq:angular-momentum-operators}
\hat L_x & = & i \left( \sin\phi \partial_\theta + \cot\theta \cos\phi \partial_\phi - \csc\theta \cos\phi \partial_\psi \right), \nn\\
\hat L_y & = & i \left( - \cos\phi \partial_\theta + \cot\theta \sin\phi \partial_\phi - \csc\theta \sin\phi \partial_\psi \right), \nn\\
\hat L_z & = & - i \partial_\phi,
\eea
which generate rotations about {\it fixed} axes. These satisfy the familiar $SO(3)$ algebra. Nevertheless, a larger algebra exist for tensorial functions on the sky~\cite{Dai:2012bc}, once three boost generators $\hat Y_a$ with $a=x,y,z$ are included. Together with $\hat L_a$, they form a Lorentz algebra,
\bea
\label{eq:lorentz-algebra}
&& \comm{\hat L_a}{\hat L_b} = i \epsilon_{abc} \hat L_c, \quad \comm{\hat L_a}{\hat Y_b} = i \epsilon_{abc} \hat Y_c, \nn\\
&& \comm{\hat Y_a}{\hat Y_b} = - i \epsilon_{abc} \hat L_c,
\eea
where $\epsilon_{abc}$ is the totally anti-symmetric Levi-Civita tensor. From the explicit form of $\hat Y_z$ given by \refeq{Yz-explicit}, we can derive the other boost generators using \refeq{lorentz-algebra}:
\bea
\label{eq:boost-operators}
\hat Y_x & = & - i \left( \sin\theta \cos\phi - \cos\theta \cos\phi \partial_\theta  + \csc\theta \sin\phi \partial_\phi \right. \nn\\
&& \left. - \cot\theta\sin\phi \partial_\psi \right), \nn\\
\hat Y_y & = & - i \left( \sin\theta \sin\phi - \cos\theta \sin\phi \partial_\theta - \csc\theta \cos\phi \partial_\phi \right.\nn\\
&& \left. + \cot\theta\cos\phi \partial_\psi \right), \nn\\
\hat Y_z & = & - i \left( \cos\theta + \sin\theta \partial_\theta \right).
\eea
Rotating the coordinate system will simply take $\hat Y_z$ to some linear combination of $\hat Y_a$'s, which generates a Lorentz boost along a different direction. Therefore, for a boost along the direction $\vect{n}$ with boost velocity $\beta$, a rapidity vector can be written $\vect{\eta}=\eta \,\vect{n}$. The general aberration kernels then read
\bea
\label{eq:kernel-mat-anyboost}
\abker{\ell'}{\ell}{m'm}{s}(\beta,\vect{n}) = \bra{s\ell' m'} e^{i\vect{\eta}\cdot \vect{\hat Y}} \ket{s\ell m},
\eea
which determines the mixing between $a'^s_{\ell' m'}$ in frame $S'$ with $a^s_{\ell m}$ in frame $S$. Since $\hat Y_a$'s are hermitian operators with respect to the inner product \refeq{inner-product}, $e^{i\eta \hat Y_z}$ and $e^{i\vect{\eta}\cdot\vect{ \hat Y}}$ are unitary, and the unitarity of the aberration kernels is thus obvious~\cite{Jeong:2013sxy}. \refeq{kernel-mat-zboost} and its generalization \refeq{kernel-mat-anyboost} are the major results of this paper.

\section{Mixing of $E$/$B$ modes}
\label{sec:EB-mixing}

The $B$-mode polarization is a unique signature in the CMB. While detecting primordial $B$ modes will be a confirmation of an inflationary background of gravitational waves, various secondary effects at late times can convert $E$ modes into $B$ modes and hence confuse the primordial signal, in particular at small scales \cite{Challinor2005, Lewis2006}. It is therefore of great importance to have accurate, unambiguous predictions for secondary $B$-mode contamination.

In this Section, we address the question of whether aberration mixes up $E$-mode and $B$-mode polarization. We find that, depending on the Doppler weight $d$ for the polarization observable, spurious $B$ modes are {\it not} produced for $d=1$, but are converted from $E$ modes for $d \neq 1$. This statement neglects cut-sky effects which we will discuss in a subsequent publication \citep{Jeong:future}.
Previously, based on leading-order expansion in $\beta$ for the polarization kernels, Ref.~\cite{Challinor:2002zh} found $E$/$B$-mode mixing for polarization observables weighted by specific intensity ($d=3$) and by frequency-integrated intensity ($d=4$), and Ref.~\cite{Amendola:2010ty} demonstrated that no mixing occurs for the case of $d=1$. Ref.~\cite{Notari:2011sb} state that they checked numerically that no mixing occurs up to $\mathcal{O}(\beta^6)$, again for $d=1$. Here, we analytically generalize to all orders in $\beta$.

The $E$/$B$-mode multipoles are related to the helical multipoles through
\bea
a^E_{\ell m} = \frac{1}{\sqrt 2} \left( a^{P_+}_{\ell m} + a^{P_-}_{\ell m} \right), \,\, a^B_{\ell m} = \frac{1}{\sqrt 2 i} \left( a^{P_+}_{\ell m} - a^{P_-}_{\ell m} \right).
\eea
Under aberration they transform as
\begin{subequations}
\bea
a^{\prime E}_{\ell' m} & = & \frac12 \sum_{\ell} \left[ \left( \abker{\ell'}{\ell}{m}{2} + \abker{\ell'}{\ell}{m}{-2} \right) a^E_{\ell m} \right. \nn\\
&& \left. + i \left( \abker{\ell'}{\ell}{m}{2} - \abker{\ell'}{\ell}{m}{-2} \right) a^B_{\ell m} \right], \\
a^{\prime B}_{\ell' m} & = & \frac12 \sum_{\ell} \left[ \left( \abker{\ell'}{\ell}{m}{2} + \abker{\ell'}{\ell}{m}{-2} \right) a^B_{\ell m} \right. \nn\\
&& \left. - i \left( \abker{\ell'}{\ell}{m}{2} - \abker{\ell'}{\ell}{m}{-2} \right) a^E_{\ell m} \right].
\eea
It can be seen that no mixing occurs if $\abker{\ell'}{\ell}{m}{2}\equiv \abker{\ell'}{\ell}{m}{-2}$. Indeed, we prove in \refapp{sign-spinweight} that
\bea
\label{eq:kernel-symmetry-s}
\abkerd{\ell'}{\ell}{m}{s}{d}(\beta) = \abkerd{\ell'}{\ell}{m}{-s}{d}(\beta)
\eea
\end{subequations}
holds for arbitrary $\beta$ {\it if and only} if $d=1$.

What value of the Doppler weight $d$ should the CMB polarization observables take? In general, it depends on which physical quantity is exactly being measured. However, if we assume that the CMB has a perfect blackbody spectrum, then a given map-making procedure should allow us to faithfully reconstruct the thermodynamic temperatures, for both linear polarizations and also for the unpolarized average, on a pixel-by-pixel basis. The map-making procedure should also produce the correct maps with the same experimental device operating in any inertial frame. In that case, independent of how the measurement is technically performed, the polarization observables have the same Doppler weight $d=1$ as the (polarization-averaged) temperature, and therefore do not have $E$ and $B$ modes mixed up under a Lorentz boost. 

In reality, foreground contaminations have spectra that differ from that of a blackbody. A non-blackbody spectrum in general will not preserve its spectrum shape under a change of reference frame. Also, no frequency-independent temperature can be unambiguously defined in the presence of foregrounds. The interpretation on the effect of Lorentz boost is then less clear than in the ideal case of blackbody spectrum. This issue, which is expected to be dependent on the details of experimental approaches and further complicates our ability to perform a "de-boosting" operation, deserves more careful consideration, but is left to a future work.

We would like to emphasize that for a single photon, a Lorentz boost does not change the direction of the polarization vector with respect to the new line of sight. In fact, the polarization plane is parallel-transported on the sky, and it only `rotates' to adjust to the curvature of the sky; there is no additional rotation about the line of sight whatsoever.

\section{Kernel recursion relations}
\label{sec:recur}
Because the harmonic-space aberration kernels are the matrix elements of a unitary transformation due to a Lorentz boost, it is reasonable to believe that one might find simple recursion relations between the matrix elements following from the algebraic properties of $\hat Y_a$'s. In this Section, we derive new, useful recursions from our operator formalism. These improve upon previous recursive algorithms~\cite{Chluba:2011zh}: (i) they follow directly from the Lorentz algebra, and are simple and elegant from a theoretical point of view; (ii) they do not rely on power expansions in $\beta$, and hence are efficient in the non-perturbative regime $\ell\gtrsim 1/\beta$; (iii) the higher spin-weight kernels are reduced to the zero spin-weight kernels in a simple way, which allows for efficient computations for polarization.

\subsection{Changing $\ell$}
As we have already seen, the $\hat Y_z$ operator does not affect $\phi$ and $\psi$. Therefore, acting $\hat Y_z$ on the state $\ket{s\ell m}$ only changes the quantum number $\ell$. 
Introducing
\begin{align}
\label{eq:C_def}
\sCml{s}{m}{\ell}&=\ell \sBml{s}{m}{\ell}=\sqrt{\frac{(\ell^2-m^2)(\ell^2-s^2)}{4\ell^2-1}}
\end{align}
for convenience, we find
\begin{align}
\label{eq:Yz-action}
i \hat Y_z \ket{s\ell m} &=\!\sCml{s}{m}{\ell+1} \ket{s\,\ell+1\,m} - \!\sCml{s}{m}{\ell} \ket{s\,\ell-1\,m}.
\end{align}
A straightforward proof of this relation is given in \refapp{Yz-action}. Next, we make use of the trivial commutator $\comm{\hat Y_z}{e^{i\eta \hat Y_z}}=0$. By taking the matrix element of both sides, we find
\bea
\bra{s\ell' m} ( \hat Y_z e^{i\eta \hat Y_z} - e^{i\eta \hat Y_z} \hat Y_z ) \ket{s\,\ell-1\, m} = 0.
\eea
Applying \refeq{Yz-action}, and also using the fact that $\hat Y_z$ is hermitian, we find a relation involving four kernels 
\begin{align}
\label{eq:kern_main_rec}
\abker{\ell'}{\ell}{m}{s} & = 
\frac{\sCml{s}{m}{\ell'}}{\sCml{s}{m}{\ell}} \abker{\ell'-1}{\,\ell-1}{m}{s}
- \frac{\sCml{s}{m}{\ell'+1}}{\sCml{s}{m}{\ell}}\abker{\ell'+1}{\,\ell-1}{m}{s} 
\nn\\
&\qquad\qquad
+ \frac{\sCml{s}{m}{\ell-1}}{\sCml{s}{m}{\ell}}  \abker{\ell'}{\ell-2}{m}{s}.
\end{align}
Thus, for $\ell>|m|$, one can compute $\abker{\ell'}{\ell}{m}{s}$ from $\abker{\ell'-1}{\,\ell-1}{m}{s}$, $\abker{\ell'}{\ell-2}{m}{s}$ and $\abker{\ell'+1}{\,\ell-1}{m}{s}$. The recursion applies to kernels with fixed $m$ and spin weight $s$. Notice that both $\ell$ and $\ell'$ change, so that the recursions remind us of the discretized version of first order partial differential equation in two dimensions.

\vspace{-2mm}
\subsection{Raising and lowering $m$}
\label{sec:raise-lower-m}
To raise and lower the azimuthal quantum number $m$, we use the operator relation \citep{Goldberg:1967spin-s} 
\bea
\label{eq:Lpm-action}
\hat L_{\pm} \ket{s\ell m} = \sqrt{(\ell \mp m)(\ell \pm m +1)} \ket{s\ell\, m \pm 1},
\eea
where $\hat L_\pm=\hat L_x\pm i \hat L_y$ are the familiar angular-momentum raising and lowering operators. Furthermore, we define for the boost generators $\hat Y_\pm=\hat Y_x \pm i \hat Y_y$. Then by combining \refeq{Yz-action}, \refeq{Lpm-action} and $\comm{\hat Y_z}{\hat L_\pm}=\pm \hat Y_\pm$, we obtain the action of $\hat Y_{\pm}$ on base states,
\begin{align}
\label{eq:Ypm-action}
\mp i \,\hat Y_{\pm} \ket{s\ell m} &= \sCml{s}{m}{\ell}\,\sqrt{\frac{\ell\mp m-1}{\ell\pm m}} \ket{s\,\ell-1\,m\pm1} 
\\ \nn
&\qquad + \sCml{s}{m}{\ell+1}\,\sqrt{\frac{\ell\pm m+2}{\ell\mp m+1}} \ket{s\,\ell+1\,m\pm 1}.
\end{align}
Using \refeq{lorentz-algebra} and the Baker-Campbell-Hausdorff formula, we can also show that
\begin{align}
\hat L_\pm\,e^{i\eta \hat Y_z} &= e^{i\eta \hat Y_z} ( \cosh \eta\,\hat L_\pm  \mp i \sinh\eta\,\hat Y_\pm ). 
\end{align}
Taking the matrix element $\bra{s\ell'\,m-1}\cdots\ket{s\ell m}$ on both sides, applying \refeqs{Lpm-action}{Ypm-action}, and also using the fact that $\hat L_\pm$ and $\hat Y_\pm$ are pairs of hermitian conjugation, respectively, we find
\bea
\label{eq:recur-raise-m}
&& \abker{\ell'}{\ell}{m}{s} = \sBml{s}{0}{\ell}\,\sinh\eta 
\sqrt{\frac{(\ell \mp m)(\ell \mp m-1)}{(\ell' \mp m)(\ell' \pm m+1)}} \abker{\ell'}{\ell-1}{m \pm 1}{s} \nn\\
&& 
\qquad+ \cosh\eta \sqrt{\frac{(\ell \mp m)(\ell \pm m+1)}{(\ell'\mp m)(\ell'\pm m+1)}} \abker{\ell'}{\ell}{m \pm 1}{s} \\
&& 
\qquad+ \sBml{s}{0}{\ell+1}\,\sinh\eta \sqrt{\frac{(\ell \pm m+1)(\ell \pm m+2)}{(\ell' \mp m)(\ell' \pm m+1)}} \abker{\ell'}{\,\ell+1}{m \pm 1}{s}. \nn
\eea
This recursively relates the kernel of $m$ to those of $m \pm 1$.

\vspace{-2mm}
\subsection{Raising and lowering $s$}
In the recursions derived so far, the spin weight $s$ has not been touched. However, it is feasible to raise and lower $s$ as well, thus relating the polarization kernels directly to those for the temperature. One realizes that there is a symmetry between the azimuthal angle $\phi$ and the roll angle $\psi$ if $(\phi,\theta,\psi)$ are interpreted as three Euler angles. While $\phi$ is associated with rotations about the $z$ axis fixed in space, $\psi$ is related to rotations about a ``body-fixed $z$'' axis -- the axis that points in the normal direction and differs from point to point on the sky.

In fact, three ``body-fixed'' angular momentum operators $\hat I_a$ with $a=x,y,z$ can be obtained by swapping $\phi$ with $\psi$ everywhere in \refeq{angular-momentum-operators},
\bea
\hat I_x & = & i \left( \sin\psi \partial_\theta + \cot\theta \cos\psi \partial_\psi - \csc\theta \cos\psi \partial_\phi \right), \nn\\
\hat I_y & = & i \left( - \cos\psi \partial_\theta + \cot\theta \sin\psi \partial_\psi - \csc\theta \sin\psi \partial_\phi \right), \nn\\
\hat I_z & = & - i \partial_\psi.
\eea
Similarly, three ``body-fixed'' boost generators $\hat Z_a$ with $a=x,y,z$ similarly follow from \refeq{boost-operators},
\bea
\hat Z_x & = & - i \left( \sin\theta \cos\psi - \cos\theta \cos\psi \partial_\theta  + \csc\theta \sin\psi \partial_\psi \right. \nn\\
&& \left. - \cot\theta\sin\psi \partial_\phi \right), \nn\\
\hat Z_y & = & - i \left( \sin\theta \sin\psi - \cos\theta \sin\psi \partial_\theta - \csc\theta \cos\psi \partial_\psi \right.\nn\\
&& \left. + \cot\theta\cos\psi \partial_\phi \right), \nn\\
\hat Z_z & = & - i \left( \cos\theta + \sin\theta \partial_\theta \right).
\eea
The symmetry between $\phi$ and $\psi$ implies that $\hat I_a$'s and $\hat Z_a$'s form another copy of Lorentz algebra,
\bea
&& \comm{\hat I_a}{\hat I_b} = i \epsilon_{abc} \hat I_c, \quad \comm{\hat I_a}{\hat Z_b} = i \epsilon_{abc} \hat Z_c, \nn\\
&& \comm{\hat Z_a}{\hat Z_b} = - i \epsilon_{abc} \hat I_c.
\eea
The spin weight $s$ is nothing but the eigenvalue of the ``body-fixed'' $\hat I_z$ operator. Because $\comm{\hat I_a}{\hat L_b}=0$, we have simultanenous eigenstates for $\hat L_z$ and $\hat I_z$,
\bea
\hat L_z \ket{s\ell m} = m \ket{s \ell m}, \quad \hat I_z \ket{s\ell m} = s \ket{s \ell m},
\eea
which establishes a formal symmetry between $m$ and $s$. This implies that we can contruct $\hat I_\pm=\hat I_x \pm i \hat I_y$ to raise and lower $s$,
\bea
\label{eq:Ipm-action}
\hat I_{\pm} \ket{s\ell m} = \sqrt{(\ell \mp s)(\ell \pm s +1)} \ket{s \pm 1\, \ell m}.
\eea
Moreover, for ``body-fixed'' boost generators we similarly define $\hat Z_\pm=\hat Z_x \pm i \hat Z_y$. The result analogous of \refeq{Ypm-action} but for $s$ thus reads
\begin{align}
\label{eq:Ipm-action}
\mp i\, \hat I_{\pm} \ket{s\ell m} &= \sCml{s}{m}{\ell}\,\sqrt{\frac{\ell\mp s-1}{\ell\pm s}} \ket{s\pm1\,\ell-1\,m}
\\ \nn
&\qquad 
+\sCml{s}{m}{\ell+1}\,\sqrt{\frac{\ell\pm s+2}{\ell\mp s+1}} \ket{s\pm1\,\ell+1\,m}.
\end{align}
Repeating a derivation similar to the one in \refsec{raise-lower-m}, and also noting that $\hat Z_z=\hat Y_z$, we can write
\begin{align}
\hat I_\pm\,e^{i\eta \hat Y_z} &= e^{i\eta \hat Y_z} ( \cosh \eta\,\hat I_\pm  \mp i \sinh\eta\,\hat Z_\pm).
\end{align}
In analogy to the $m$-raising case, we take the matrix elements $\bra{s-1\,\ell' m}\cdots\ket{s\ell m}$ on both sides, and obtain a recursion similar to \refeq{recur-raise-m}, with the roles of $m$ and $s$ exchanged:
\bea
&& \abker{\ell'}{\ell}{m}{s} = 
\sBml{0}{m}{\ell}\,
\sinh\eta \sqrt{\frac{(\ell \mp s)(\ell \mp s-1)}{(\ell' \mp s)(\ell' \pm s+1)}} \abker{\ell'}{\ell-1}{m}{s \pm 1} \nn\\
&& 
\label{eq:rec_spin_raise}
\qquad+ \cosh\eta \sqrt{\frac{(\ell \mp s)(\ell \pm s+1)}{(\ell' \mp s)(\ell' \pm s+1)}} \abker{\ell'}{\ell}{m}{s \pm 1} \\
&& 
\qquad+ \sBml{0}{m}{\ell+1}\,
\sinh\eta \sqrt{\frac{(\ell \pm s+1)(\ell \pm s+2)}{(\ell' \mp s)(\ell' \pm s+1)}} \abker{\ell'}{\ell+1}{m}{s \pm 1}. \nn
\eea
This recursion relates the kernels of spin weight $s$ to those of spin-weight raised/lowered by one unit. As we will see below, this expression implies that the temperature and polarization kernels are very similar once $\ell\gg s$.

\subsection{A practical recursive scheme}
To make practical use of the recursions given above, a few additional steps are required. First of all, we have to decide how to run through the recursions, combining them in a convenient way to a numerically stable scheme. The procedure in particular depends on the required initial conditions that can be obtained in a simple (closed) form. Secondly, we want to compute the kernel elements in the most economic way, making use of its symmetries.

For the temperature kernel, a method based on term-by-term expansions in $\beta$ was already given by \cite{Chluba:2011zh}. Once the temperature kernel ($s=0$) is computed, by applying the $s$-raising operator, \refeq{rec_spin_raise}, twice the required polarization kernel is directly obtained and we are done. To compute the temperature kernel, because $\abker{\ell'}{\ell}{m}{0}=\abker{\ell'}{\ell}{-m}{0}$ and $\abker{\ell'}{\ell}{m}{0}=(-1)^{\ell+\ell'}\!\abker{\ell}{\ell'}{m}{0}$, we only need those elements for $m\geq 0$ and $\ell\leq \ell'$, reducing the number of independent coefficients by a factor of $\simeq 4$.
For our purposes, this method in principle is sufficient, however, with the expressions given above we can simplify the computation significantly, as we explain now. 

\subsubsection{Applying the recursions}

As shown earlier \citep{Chluba:2011zh}, at $\ell\gtrsim 1/\beta$ the kernel widens, coupling more and more neighboring $\ell$-modes. In principle, by knowing all matrix elements $\abker{\ell}{\ell}{m}{0}$ (i.e., the diagonal at fixed $m$) for $\ell\leq \ell_{\rm max}$ and using $\abker{\ell'}{m}{m}{0}=0$ for $\ell'<m$, one could obtain all elements $\abker{\ell+1}{\ell-1}{m}{0}$, $\abker{\ell+2}{\ell-2}{m}{0}$, $\abker{\ell+3}{\ell-3}{m}{0}$ etc. using Eq.~\eqref{eq:kern_main_rec}. Similarly, those elements $\abker{\ell+2}{\ell-1}{m}{0}$, $\abker{\ell+3}{\ell-2}{m}{0}$, $\abker{\ell+4}{\ell-3}{m}{0}$ etc. could be obtained by knowing $\abker{\ell+1}{\ell}{m}{0}$ (i.e., the first off-diagonal). In this way, one could nicely compute all kernel elements $\abker{\ell'}{\ell}{m}{0}$ for $\ell'-\ell\leq 2 \ell_{\rm max}$. Since the off-diagonal kernel elements drop like $\Kerns{0}{\ell}{m}{\ell'}\simeq \beta^{|\ell'-\ell|}$ in amplitude \citep{Chluba:2011zh}, one could stop the recursions at some finite value of $\Delta \ell = \ell'-\ell$ obtaining an extremely economic method for computing the aberration kernel. We were, however, unable to find a simple way to give all the required initial conditions, $\abker{\ell}{\ell}{m}{0}$ and $\abker{\ell+1}{\ell}{m}{0}$, so that this procedure is impractical.

Instead we start our recursions at $\ell=\ell'=m=0$, using $\abker{0}{0}{0}{0}=\eta/[\beta \gamma]$. We then apply the term-by-term expansion in $\beta$ given by \cite{Chluba:2011zh} to obtain the elements $\abker{\ell'}{0}{0}{0}$ for $0\leq \ell'\leq 2\ell_{\rm max}$. Afterwards, we apply the $\ell$-changing recursion, Eq.~\eqref{eq:kern_main_rec}, to fill in the remaining matrix elements for $\abker{\ell'}{\ell}{0}{0}$ up to $\ell+\ell'\leq 2\ell_{\rm max}$, preceding in a row-by-row manner, fixing $\ell$ and changing $\ell'$ within the row. By applying the $m$-raising operator, Eq.~\eqref{eq:recur-raise-m}, we then compute the row $\abker{\ell'}{1}{1}{0}$ for $\ell'\leq 2\ell_{\rm max}-1$ from which we obtain the whole layer $\abker{\ell'}{\ell}{1}{0}$ for $\ell+\ell'\leq 2\ell_{\rm max}-1$, applying Eq.~\eqref{eq:kern_main_rec} again. We continue this procedure until $\ell=\ell'=m=\ell_{\rm max}$. This scheme works very well after rewriting the recursions, as we explain below.

\vspace{2mm}
\subsubsection{Initial conditions and recursion for $\Kerns{0}{0}{0}{\ell'}$}
To start the computation, we need to provide the initial conditions and recursion for $\Kerns{0}{0}{0}{\ell'}$. As shown by \cite{Chluba:2011zh}, for the temperature kernel element $\Kerns{0}{m}{m}{m}(\beta)$ we have
\begin{align}
\Kerns{0}{m}{m}{m}(\beta)
&=\frac{1}{\gamma^{m+1}}\sum_{k=0}\frac{(2k+m)!}{2^k k! \,m!} \frac{(2m+1)!! \, \beta^{2k}}{(2m+2k+1)!!} 
\\ \nn
&={}_2F_1\left(\frac{m}{2}+\frac{1}{2},\frac{m}{2}+1, m+\frac{3}{2},\beta^2\right)/\gamma^{m+1},
\end{align}
where ${}_2F_1\left(a, b, c, z\right)$ is the hypergeometric function. We generally use this expression for all matrix elements $\Kerns{0}{m}{m}{m}(\beta)$, even if in principle for $m>0$ simple $m$-raising would work. To obtain all the matrix elements $\Kerns{0}{0}{0}{\ell'}$ for $0\leq\ell' \leq 2\ell_{\rm max}$ we need to precede in a term-by-term manner as cancelations of terms prevent the direct recursions from converging. From the results of \cite{Chluba:2011zh}, we find 
\begin{align}
\Kerns{0}{0}{0}{\ell'}(\beta)
&=\frac{\beta^{\ell'}}{2^{\ell'}\! \gamma}\sum_{k=0}\,\kappa^{\ell'}_{k},
\qquad \kappa^{0}_{k}=\frac{\beta^{2k}}{2k+1}
\\ \nn
\kappa^{\ell'}_{k}&= \frac{2\ell'}{\sqrt{4{\ell'}^2-1}} \kappa^{\ell'-1}_{k} +\frac{(\ell'+1)\beta^2}{2\sqrt{4(\ell'+1)^2-1}} \kappa^{\ell'+1}_{k-1}.
\end{align}
We scaled out the main term $\simeq (\beta/2)^{\ell'}/\gamma$ which makes all $\kappa^{\ell'}_{0}$ become of order unity at large $\ell'$. For $\beta \lesssim 0.01$ and $\ell_{\rm max}\lesssim 4000$ we never needed more than $128$ terms in the expansion of $\beta$. For better convergence, we furthermore used long double precision in the computations.

\begin{widetext}

\subsubsection{Rewriting the recursions}
\label{sec:rewrite_rec}
The off-diagonal kernel elements drop like $\Kerns{0}{\ell}{m}{\ell'}\simeq \beta^{|\ell'-\ell|}$ in amplitude \citep{Chluba:2011zh}. This means that for large $\Delta \ell = \ell'-\ell$, the kernel elements become extremely small, and since in the computation elements for $\ell\ll \ell'\simeq 2\ell_{\rm max}$ are needed, it is crucial to rewrite the kernel recursions to improve the numerical stability. For this one has to scale out the leading order behavior of the kernel.
Applying the boost operator $i\hat{Y}_z$ several times to the state $\left| s\ell m\right>$ and then projecting onto $\left| s\ell' m\right>$, with $\Kern{\ell'}{\ell}=\left<s\ell'm\right| \exp(i\eta \hat{Y}_z)\left|s\ell m\right>$ shows that the leading order term of the kernel scales like
\bea
\label{eq:leading_order}
\Kern{\ell+\Delta\ell}{\ell}\approx \frac{\eta^{\Delta \ell}}{\Delta \ell !}\,\prod_{k=1}^{\Delta\ell} \Cfunc{\ell+k} 
=\frac{\eta^{\Delta \ell}}{\Delta \ell !}\,\frac{(2\ell-1)!!}{(2\ell'-1)!!} \sqrt{\frac{(2\ell+1)}{(2\ell'+1)}\frac{(\ell'+s)!}{(\ell+s)!}\frac{(\ell'-s)!}{(\ell-s)!}\frac{(\ell'+m)!}{(\ell+m)!}\frac{(\ell'-m)!}{(\ell-m)!}}
\stackrel{\stackrel{\ell\gg s, m}{\downarrow}}{\approx} \frac{\eta^{\Delta \ell}}{2^{\Delta \ell} \Delta \ell !}\,\frac{\ell'!}{\ell!} 
\eea
for $\Delta \ell=\ell'-\ell\geq 0$ and $\eta=\ln([1+\beta]/[1-\beta])/2\approx \beta$. Rescaling the kernel by the leading order term, introducing $\Kernb{\ell}{\ell'}=\Kern{\ell'}{\ell}/[\frac{\eta^{\ell'-\ell}}{(\ell'-\ell)!}\,\prod_{k=1}^{\ell'-\ell} \Cfunc{\ell+k}]$ and using Eq.~\eqref{eq:kern_main_rec} we find
\begin{align}
\Kernb{\ell}{\ell'}
&=\Kernb{\ell-1}{\ell'-1}
-\frac{(\Cfunc{\ell'+1})^2\,\eta^2\,\Kernb{\ell-1}{\ell'+1}}{(\ell'-\ell+1)(\ell'-\ell+2)}
+\frac{(\Cfunc{\ell-1})^2\,\eta^2\,\Kernb{\ell-2}{\ell'}}{(\ell'-\ell+1)(\ell'-\ell+2)}
\end{align}
for $\ell'\geq \ell$. The matrix elements $\Kernb{\ell}{\ell'}$ are now all of order unity and hence the new recursion is numerically more stable. In a similar manner, we obtain
\begin{align}
\Kernb{\ell}{\ell'}
&=\frac{\ell+m}{\ell'+m}
\left[\cosh \eta \,\Kernsb{s}{\ell}{m-1}{\ell'}
+\frac{\sinh \eta}{\eta}\left(\frac{\ell'-\ell}{\ell+m}\,\Kernsb{s}{\ell+1}{m-1}{\ell'}
+(\eta \sCml{s}{0}{\ell})^2\,\frac{\ell+m-1}{\ell'-\ell+1}\,\Kernsb{s}{\ell-1}{m-1}{\ell'}
\right)
\right]
\nn\\[2mm]
\Kernb{\ell}{\ell'}
&=\frac{\ell+s}{\ell'+s}
\left[\cosh \eta \,\Kernsb{s-1}{\ell}{m}{\ell'}
+\frac{\sinh \eta}{\eta}\left(\frac{\ell'-\ell}{\ell+s}\,\Kernsb{s-1}{\ell+1}{m}{\ell'}
+(\eta \sCml{0}{m}{\ell})^2\,\frac{\ell+s-1}{\ell'-\ell+1}\,\Kernsb{s-1}{\ell-1}{m}{\ell'}
\right)
\right].
\end{align}
for the $m$ and $s$-raising recursions. These expressions are at the core of our numerical recursion scheme. We find them to work even up to $\beta\simeq 0.01$. In this case, the kernel is already rather broad at large $\ell$, reaching $\Delta \ell \simeq 70$ at $\ell\simeq 4000$. For $\beta\simeq 10^{-3}$ we find $\Delta \ell=10-20$ to suffice.

\end{widetext}

\begin{figure}
\centering
\includegraphics[width=\columnwidth]{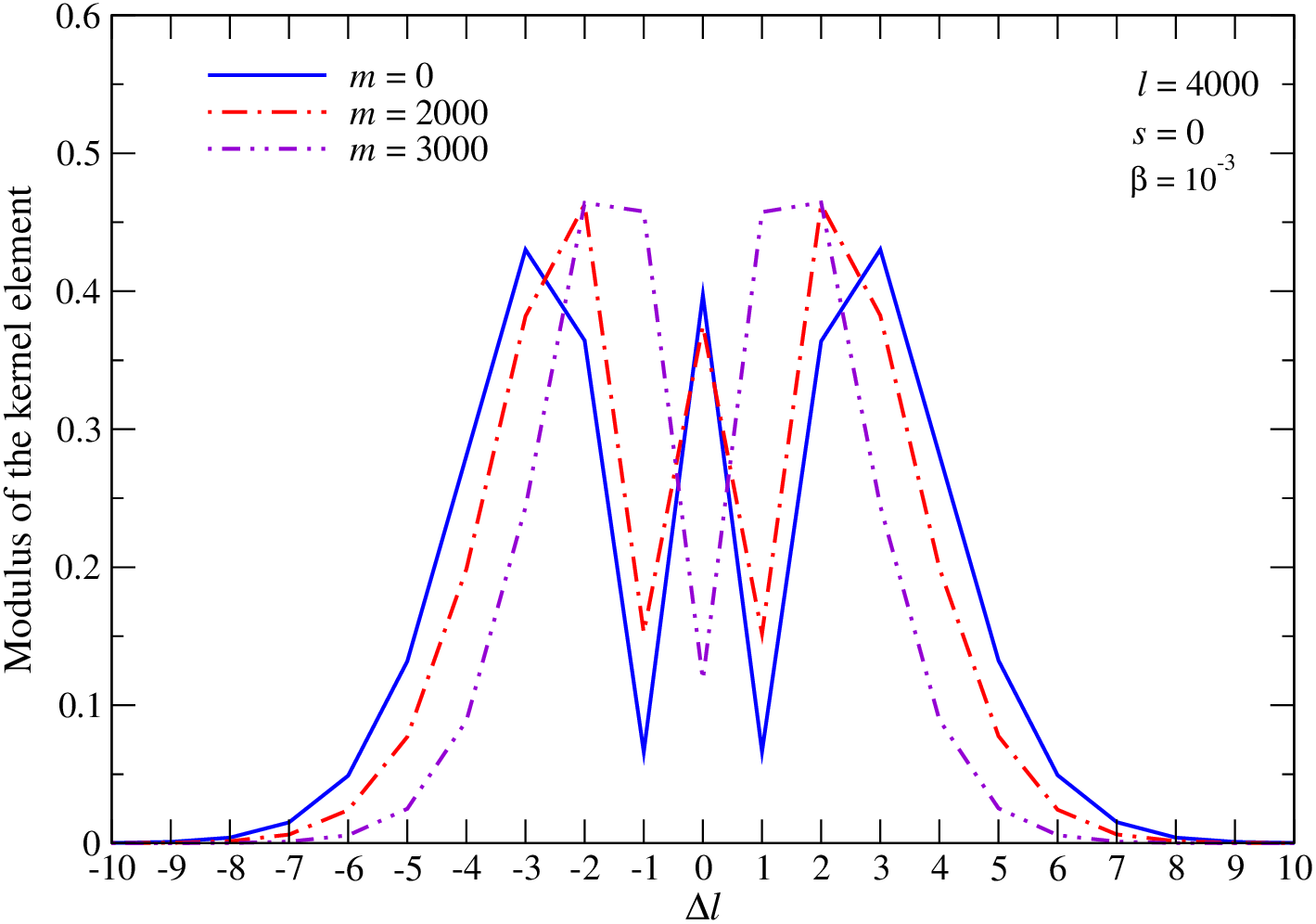}
\\[5mm]
\includegraphics[width=\columnwidth]{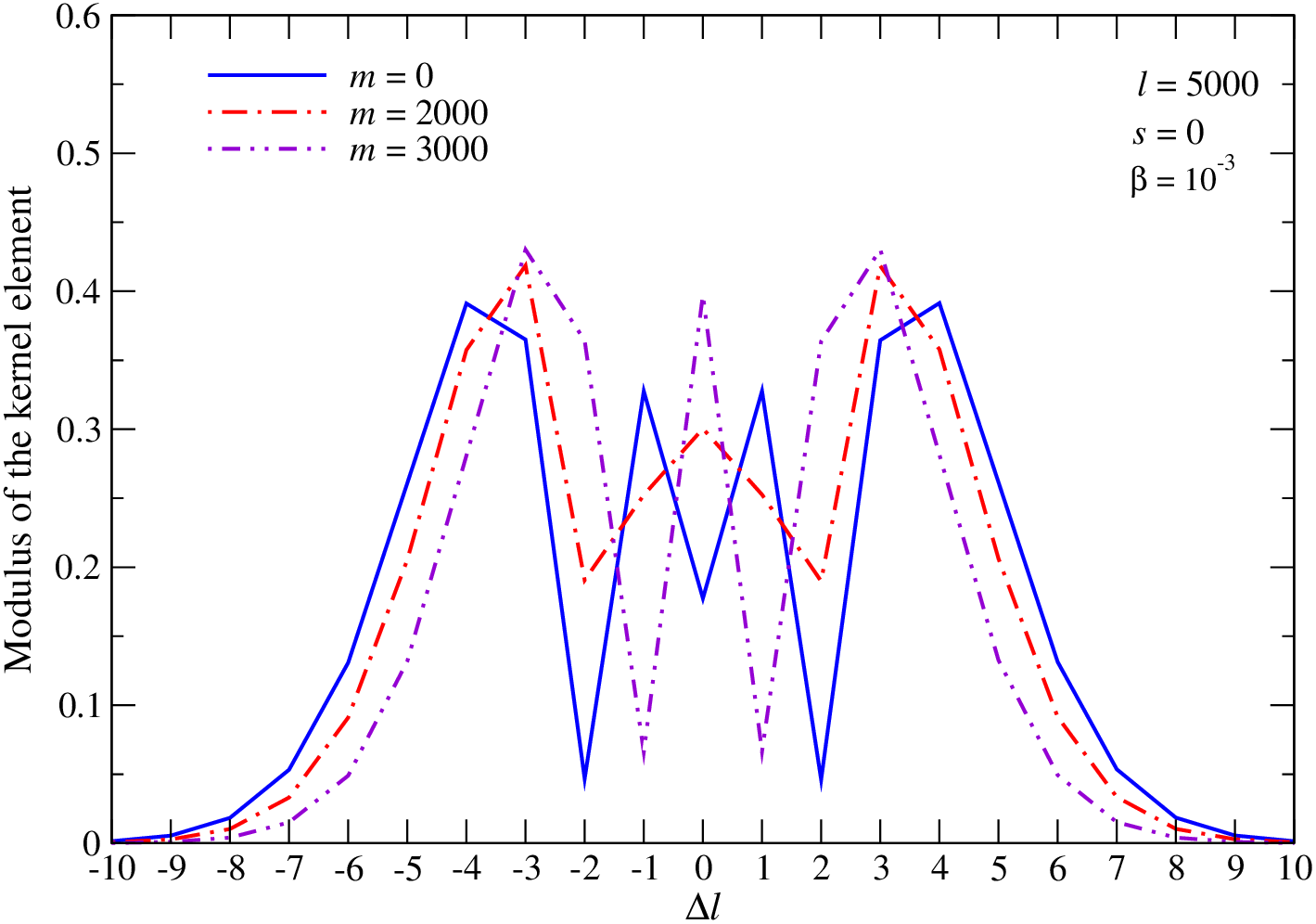}
\caption{Modulus of the temperature kernel $\Kerns{0}{\ell}{m}{\ell+\Delta \ell\,}$ for $\beta=10^{-3}$, $\ell=4000$ and $\ell=5000$. For both cases, the corresponding polarization kernel ($s=2$) is extremely similar.}
\label{fig:examples}
\end{figure}
\section{Differential equation representation}
\label{sec:ode-rep}
%
With the operator representation, we can write a system of coupled ordinary differential equations (ODEs) for the aberration kernels. Using the definition of the aberration kernel element, $\Kerns{s}{\ell}{m}{\ell'}=\left<s\ell'm\right| \exp(i\eta \hat{Y}_z)\left|s\ell m\right>$, gives
\begin{align}
\partial_\eta \Kerns{s}{\ell}{m}{\ell'}
&=\left<s\ell'm\right| i \hat{Y}_z \exp(i\eta \hat{Y}_z)\left|s\ell m\right>
\nn\\[2mm]
&=\Cfunc{\ell+1}\Kerns{s}{\ell+1}{m}{\ell'} -  \Cfunc{\ell}\Kerns{s}{\ell-1}{m}{\ell'}
\nn\\[2mm]
\label{eq:ODE_appro}
&=\Cfunc{\ell'+1}\Kerns{s}{\ell}{m}{\ell'+1} -  \Cfunc{\ell'}\Kerns{s}{\ell}{m}{\ell'-1}.
\end{align}
Notice that the two independent ways of computing the $\eta$-derivative also directly give the recursion \refeq{kern_main_rec}.
For $\eta = 0$, we have the initial condition $\Kerns{s}{\ell}{m}{\ell'}=\delta_{\ell \ell'}$. It is furthermore clear that for finite $\eta$ the kernel only attains non-zero values in a limited range $|\ell' -\ell| < \Delta \ell$. We can thus write a system of ODEs in some finite range around the diagonal elements $\ell=\ell'$ for each $m$ (setting the matrix elements at the boundaries to zero), and then solve it as a function of $\eta$. The system is rather sparse and an explicit Runge-Kutta scheme turns out to be sufficient for solving it. We successfully used a Runge-Kutta-Fehlberg method with adaptive step size control.

The ODE representation has several benefits over the recursion scheme. First of all, it works for any spin weight without having to worry about specific initial conditions. It also does not matter if the value of $\beta$ is large or small (the integration takes a little longer for larger $\beta$). In contrast to the recursion scheme, to obtain kernel elements for large $\ell, \ell'$, in the ODE approach it is furthermore unnecessary to compute all elements up to these values. Finally, the workload is significantly reduced, since generally only matrix elements for small $\Delta \ell$ are required. 

Our final ODE scheme takes about $\simeq 80$ seconds to compute all non-negligible kernel elements for $\beta=10^{-3}$ and $\ell_{\rm max} \simeq 4000$ on a single core (standard laptop). Parallelization of the computation is straightforward and scales very well, while this is more complicated for the recursion method.
For comparison, our best recursion scheme takes about $\simeq 35$ min for the same computation, while direct integration methods remain impractical. This large increase in the performance provides the basis for full sampling over different values of $\beta$. A few examples computed with our ODE scheme for $\beta=10^{-3}$ and large $\ell$ are given in Fig.~\ref{fig:examples}. 

For large $\ell$, both temperature and polarization kernels coincide to high precision, so that we only show the curves for $s=0$. This is not surprising at the relative level of $\simeq \mathcal{O}(s/\ell)$, however, it turns out that the difference is even smaller comparable to $\simeq \mathcal{O}(s\Delta \ell/\ell^2, \eta^2)$.
To understand this aspect a little better, let us rewrite Eq.~\eqref{eq:rec_spin_raise} as 
\begin{align}
\label{eq:rec_spin_raise_rew}
\abker{\ell'}{\ell}{m}{s} &= 
 \sqrt{\frac{(\ell \mp s)(\ell \pm s+1)}{(\ell' \mp s)(\ell' \pm s+1)}} 
 \left[\frac{\sCml{s\pm1}{m}{\ell}\,\sinh\eta}{\ell\pm s+1}\!
\abker{\ell'}{\ell-1}{m}{s \pm 1}
\right.
\\ 
&\qquad\quad + 
\left.\cosh\eta \abker{\ell'}{\ell}{m}{s \pm 1} 
+ \,\frac{\sCml{s\pm1}{m}{\ell+1}\,\sinh\eta}{\ell\mp s} \! \abker{\ell'}{\ell+1}{m}{s \pm 1}
\right].
\nn
\end{align}
This expressions shows that one modulation of the kernel values is caused by the difference between $\ell$ and $\ell'$ which is captured by an overall normalization coefficient. For small $\Delta \ell$, this gives $\sqrt{(\ell \mp s)(\ell \pm s+1)/[(\ell' \mp s)(\ell' \pm s+1)]} \approx 1-\Delta \ell / \ell +\mathcal{O}(\Delta \ell^2 / \ell^2)$. However, at lowest order in $\eta$, this modulation is precisely canceled by the variation of the other terms, so that the overall correction is of second order. For small $\eta$ and $\ell'>\ell$, the last two terms in Eq.~\eqref{eq:rec_spin_raise_rew} are dominant and by using Eq.~\eqref{eq:ODE_appro} we find
\begin{align}
 \abker{\ell'}{\ell}{m}{s} &\approx   
\left[1-\frac{\Delta\ell}{\ell}+\frac{\eta \,\partial_\eta}{\ell\mp s}\right]
\abker{\ell'}{\ell}{m}{s \pm 1}.
\nn 
\end{align}
Thus, with $\eta \,\partial_\eta\abker{\ell'}{\ell}{m}{s \pm 1}\approx \Delta \ell [1+\mathcal{O}(\eta^2)] \abker{\ell'}{\ell}{m}{s \pm 1}$, this implies $\abker{\ell'}{\ell}{m}{s}\approx \left(1\pm \frac{s}{\ell}\frac{\Delta \ell}{\ell}\right)\abker{\ell'}{\ell}{m}{s \pm 1}$ and $\abker{\ell'}{\ell}{m}{2}\approx \left(1\pm \frac{3\Delta \ell}{\ell^2}\right)\abker{\ell'}{\ell}{m}{0}$, confirming our statement. 

For similar reasons, changes of the magnetic quantum number $m\ll \ell$ will cause corrections to the kernel of order $\mathcal{O}(m \Delta \ell /\ell^2)$.

\subsection{Asymptotic expressions for the kernel}
\label{sec:asymp_exp}
\begin{figure}
\centering
\includegraphics[width=\columnwidth]{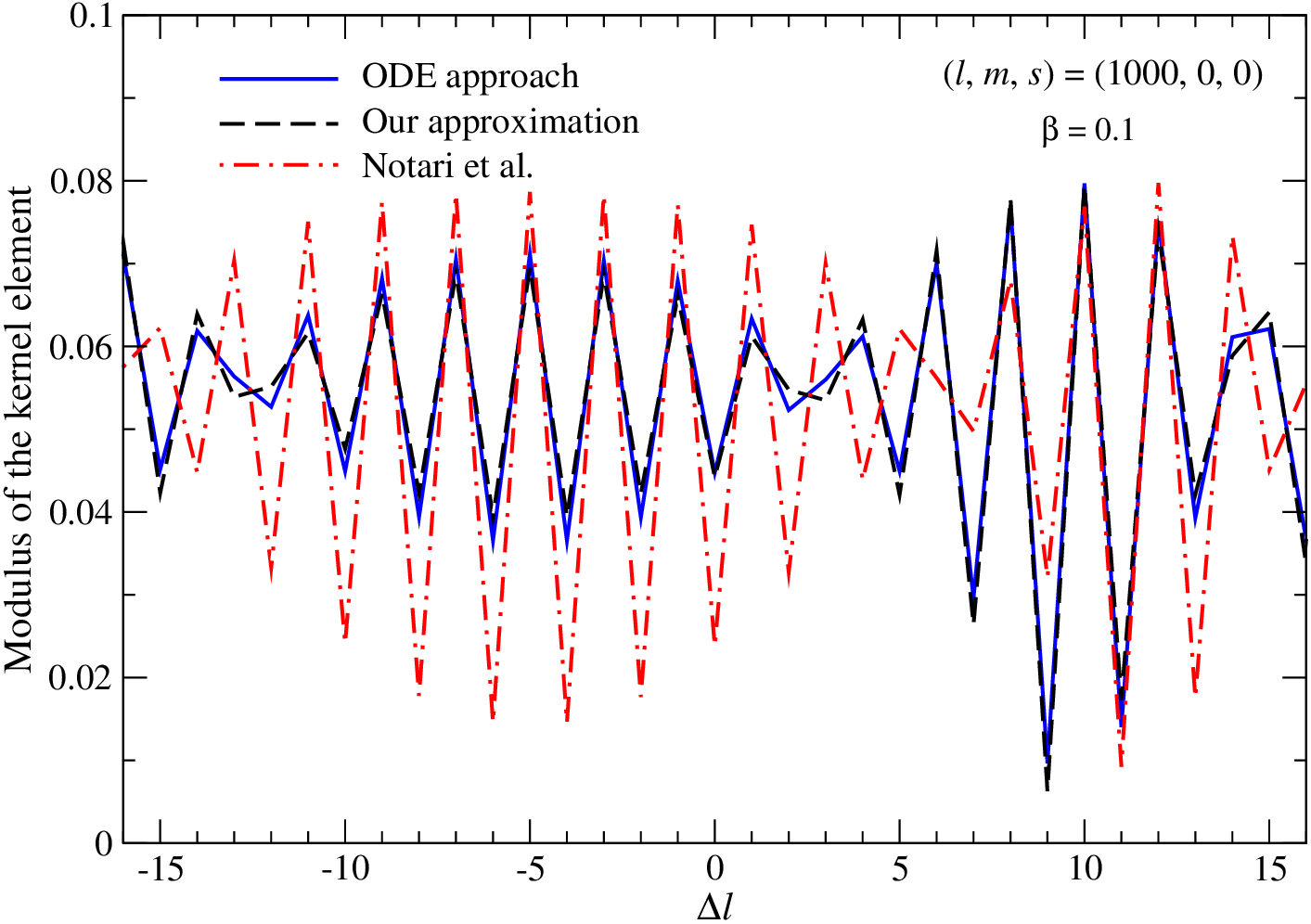}
\caption{Modulus of $\Kerns{0}{\ell}{m}{\ell+\Delta \ell\,}$ at $\ell\simeq \ell'$ for $\ell=1000$, $m=0$ and $\beta=0.1$. We compare our numerical result with the approximation Eq.~\eqref{eq:K_approx_II} and Eq.~(8)-(10) given in \cite{Notari:2011sb}.}
\label{fig:examples_large}
\end{figure}

From Eq.~\eqref{eq:ODE_appro}, we can also obtain asymptotic expressions for the aberration kernel in the limit of large $\ell$ and $\ell'$. Introducing the new variable $\eta_\ell=\Cfunc{\ell}\eta$ we find
\begin{align}
\partial_{\eta_\ell} \Kerns{s}{\ell}{m}{\ell'}
&=\frac{\Cfunc{\ell+1}}{\Cfunc{\ell}}\Kerns{s}{\ell+1}{m}{\ell'} - \Kerns{s}{\ell-1}{m}{\ell'}
\stackrel{\stackrel{\ell\gg 1}{\downarrow}}{\approx}   \Kerns{s}{\ell+1}{m}{\ell'}-\Kerns{s}{\ell-1}{m}{\ell'}.
\nn
\end{align}
The last line can be identified with the recurrence relation $2\partial_x J_n(x)=J_{n-1}(x)-J_{n+1}(x)$ for the Bessel function of first kind, $J_n(x)$, when setting $x\equiv 2 \eta_\ell$ and $n= \ell'-\ell=\Delta\ell$. Thus
\begin{align}
\Kerns{s}{\ell}{m}{\ell'}(\eta)
&\stackrel{\stackrel{\ell\gg 1, |\Delta \ell|}{\downarrow}}{\approx} J_{\Delta\ell}\left(2 \eta \, \Cfunc{\ell}\right)
\label{eq:K_approx}
\stackrel{\stackrel{\ell\gg 1, s, |\Delta \ell| }{\downarrow}}{\approx}
J_{\Delta\ell}\left(\eta \, \sqrt{\ell^2-m^2} \right).
\end{align}
We find that this expression already works very well for large $\ell$ as long as the kernel does not become too wide so that the assumption $|\Delta\ell |\ll \ell$ breaks down. Since to leading order in $x$ we have $J_n(x)\simeq x^n/[2^n n!]$, by comparing with the leading order term of $\Kerns{s}{\ell}{m}{\ell'}(\eta)$, Eq.~\eqref{eq:leading_order}, we can further improve the approximation:
\begin{align}
\Kerns{s}{\ell}{m}{\ell'}(\eta)
&\stackrel{\stackrel{\ell\gg 1, |\Delta \ell|}{\downarrow}}{\approx} 
J_{\Delta \ell}\left(2 \eta \left[\prod_{k=1}^{\Delta \ell} \Cfunc{\ell+k}\right]^{\frac{1}{\Delta \ell}}\right)
\label{eq:K_approx_II}
\nn\\[2mm]
&\,\,\,\,\,\,\approx 
J_{\Delta \ell}\left(\eta \left[\frac{(\ell'+m)!(\ell'-m)!}{(\ell+m)!(\ell-m)!}\right]^{\frac{1}{2\Delta \ell}}\right)
\end{align}
for $\ell' > \ell$, and then use $\Kerns{s}{\ell}{m}{\ell}(\eta)\approx J_{0}\left(\eta \, \sqrt{\ell^2-m^2} \right)$ and $\Kerns{s}{\ell}{m}{\ell'}(\eta)=(-1)^{\ell'-\ell}\Kerns{s}{\ell'}{m}{\ell}(\eta)$ otherwise.
This is similar to the expressions in Eq.~(8)-(10) given of \cite{Notari:2011sb}, however, there the functional form was obtained from fits to the numerical results at $\ell\lesssim 700$ rather than by analytic arguments. Our expression also works well for very large values of $\beta$. This is illustrated in Fig.~\ref{fig:examples_large} for $\beta=0.1$, $\ell=1000$ and $m=0$. Even for these extreme values of $\beta$, our approximation reproduces the main trend and amplitude of the numerical result, while Eq.~(8)-(10) of \cite{Notari:2011sb} become more crude \footnote{In a footnote, Notari et al \cite{Notari:2011sb} suggested the replacement $\beta\rightarrow \beta/(1-\beta^2)^{1/4}$ to improve the convergence of their expressions. While the leading order correction $\beta/(1-\beta^2)^{1/4}\simeq \beta+\beta^3/4$ differs from $\eta \simeq \beta -\beta^3/3$, we find that this indeed does improve the validity of the approximations provided there.}. Still, the approximation Eq.~\eqref{eq:K_approx_II} is valid only at $\Delta \ell/\ell \lesssim 1$, and since the kernel becomes wide as $\ell$ and $\beta$ increase \citep{Chluba:2011zh}, the applicability of the Bessel approximation is generally limited.

We carefully checked the precision of the approximations against the results obtained with the ODE approach and found that overall the typical error is very small ($\simeq 0.1\%-5\%$ for $\beta =10^{-3}$ and $\ell \leq 4000$). However, even for rather small $\Delta \ell \simeq 1$, $\ell\gg 1$ and $\beta\simeq 10^{-3}$ we occasionally find that the approximation can be off by a large amount, when the kernel value is close to zero-crossing (e.g., for $\beta=10^{-3}$, $\ell\simeq \ell'\simeq 2404$ and $m\simeq 0$, which is off by a factor of $\simeq 1.5$). Also, the approximation is generally less accurate for $\ell\simeq m$. We thus do not recommend using the expressions for real computations, also because the ODE approach already is very fast and reliable.

\vspace{10mm}
\subsection{Series expansion in orders of $\eta$}
\label{sec:series_exp}
From Eq.~\eqref{eq:ODE_appro}, we can also obtain simple term-by-term approximations for the aberration kernels. Rescaling them by $N_{\ell' \ell}=\eta^{\Delta \ell} \prod_{k=1}^{\Delta \ell} \Cfunc{\ell+k}$, for $\ell' > \ell$ Eq.~\eqref{eq:ODE_appro} becomes
\begin{align}
\Delta \ell\,\Kernsb{s}{\ell}{m}{\ell'}+\eta \,\partial_{\eta} \Kernsb{s}{\ell}{m}{\ell'}
&=\Kernsb{s}{\ell+1}{m}{\ell'}- (\eta \Cfunc{\ell})^2 \Kernsb{s}{\ell-1}{m}{\ell'},
\nn
\end{align}
where $\Kernsb{s}{\ell}{m}{\ell'}=\Kerns{s}{\ell}{m}{\ell'}/N_{\ell' \ell}$. Inserting the series ansatz $\Kernsb{s}{\ell}{m}{\ell'}=\sum_{k=0}^\infty (-1)^k \abkappab{\ell'}{\ell}{m}{s}{k} \eta^{2k}/(2k+\Delta \ell)!$, after collecting terms we find
\begin{align}
\abkappab{\ell'}{\ell}{m}{s}{k}
&=\abkappab{\ell'}{\ell+1}{m}{s}{k} + (\Cfunc{\ell})^2 \abkappab{\ell'}{\ell-1}{m}{s}{k-1},
\label{eq:first_rec_kappa}
\end{align}
with $\abkappab{\ell'}{\ell}{m}{s}{0}=1$ and $\abkappab{\ell'}{\ell}{m}{s}{k}=0$ for $k<0$.

For $\ell=\ell'$, we proceed similarly, finding
\begin{align}
\abkappab{\ell}{\ell}{m}{s}{k}
&=(\Cfunc{\ell+1})^2\abkappab{\ell+1}{\ell}{m}{s}{k-1} + (\Cfunc{\ell})^2 \abkappab{\ell}{\ell-1}{m}{s}{k-1}.
\label{eq:second_rec_kappa}
\end{align}
With $\abkappab{\ell'}{\ell}{m}{s}{0}=1$, for the $\eta^2$ correction to the diagonal term this equation directly implies 
\begin{align}
\abkappab{\ell}{\ell}{m}{s}{1}=(\Cfunc{\ell})^2+(\Cfunc{\ell+1})^2.
\end{align}
Inserting this back into Eq.~\eqref{eq:first_rec_kappa}, we then find
\begin{align}
\label{eq:kappa_1_llp_sol}
\abkappab{\ell'}{\ell}{m}{s}{1}=\sum_{k=0}^{\Delta \ell+1}(\Cfunc{\ell+k})^2.
\end{align}
for $\ell' > \ell$. Repeating the process, we have
\begin{subequations}
\begin{align}
\abkappab{\ell}{\ell}{m}{s}{2}&=\sum_{k=0}^{1}\sum_{p=0}^{2}(\Cfunc{\ell+k})^2(\Cfunc{\ell+k+p-1})^2
\\
\abkappab{\ell'}{\ell}{m}{s}{2}&=\sum_{k=0}^{1}\sum_{p=0}^{2}(\Cfunc{\ell'+k})^2(\Cfunc{\ell'+k+p-1})^2
\nn \\&\qquad 
+ \sum_{k=0}^{\Delta \ell-1}  \sum_{p=0}^{\Delta \ell+2-k} (\Cfunc{\ell+k})^2  (\Cfunc{\ell-1+k+p})^2
\end{align}
\end{subequations}
for the $\eta^4$ correction to the kernels. Higher order terms can be obtained in a similar way, but generally it is simpler to just evaluate the recursions Eq.~\eqref{eq:first_rec_kappa} and \eqref{eq:second_rec_kappa} in an alternating manner, so that we do not give additional explicit expressions here. We note that the form of the recursions also explicitly shows that the kernel for $d=1$ does not depend on the sign of the spin weight and hence directly proves $\Kerns{s}{\ell}{m}{\ell'}(\eta)\equiv \Kerns{-s}{\ell}{m}{\ell'}(\eta)$ required to avoid $E$/$B$-mode mixing (Sect.~\ref{sec:EB-mixing}). 

From Eq.~\eqref{eq:kappa_1_llp_sol}, we can also understand why the approximation Eq.~\eqref{eq:K_approx_II} is only expected to work for $\Delta \ell\ll \ell$. The first two terms in the Taylor series are
\begin{align}
\Kerns{s}{\ell}{m}{\ell'}(\eta)
&\approx \eta^{\Delta \ell} \prod_{k=1}^{\Delta \ell} \Cfunc{\ell+k}
\left[\frac{1}{\Delta \ell!}-\frac{\sum_{k=0}^{\Delta \ell+1}(\Cfunc{\ell+k} \eta)^2}{(\Delta \ell+2)!}\right]
\nn\\\nn
&\approx \frac{(2\Cfunc{\ell}\,\eta)^{\Delta \ell}}{2^{\Delta\ell}\Delta \ell!}
\left[1-\frac{(2\Cfunc{\ell}\eta)^2}{4(\Delta \ell+1)}\right]
\approx J_{\Delta \ell}(2\Cfunc{\ell}\,\eta),
\end{align}
which only for $\Delta \ell\ll \ell$ and $\abkappab{\ell'}{\ell}{m}{s}{1}\approx (\Delta\ell+2)(\Cfunc{\ell})^2$ can be identified with the Bessel approximation given above.

\vspace{2mm}
\section{Conclusion}
\label{sec:concl}
In this paper, we found a novel matrix representation for the harmonic-space aberration kernels. Several useful and exact relations are then derived by utilizing the commutation relations for the rotation and boost operators. CMB observables with Doppler weight $d=1$ (e.g., the polarization-averaged temperature and the temperature-weighted Stokes parameters), have the simplest transformation properties. We showed that the $d=1$ kernels are the matrix elements of a boost operator, parameterized by the additive rapidity parameter, between two spherical harmonic base states. 

The unitarity of the boost operator leads to power conservation laws under aberration, which are valid for $d=1$. The Lorentz algebra, satisfied by generators of rotations (both space-fixed and body-fixed) and boosts, lead to recursion relations that raise or lower the spherical harmonic quantum numbers $\ell$ and $m$, or the spin weight $s$ by one unit. These provide useful identities in analytical calculations. Applying these recursions repeatedly, starting from known kernels of the lowest $\ell$, $|m|$ and $|s|$ as suitable boundary conditions, yields kernels with arbitrary $\ell$, $m$ and $s$. Based on this, the new recursion scheme developed here greatly simplifies previous recursive algorithms at both conceptual and technical levels. It also provides exact values for the aberration kernels to benchmark the accuracy of existing fitting formula.

We proved that aberration does not mix up $E$ and $B$ modes for $d=1$ polarization observables to all orders in $\beta$. We argued that for perfect blackbody spectra, $d=1$ kernels are the relevant ones for the study of CMB aberration, independent of experimental details. In the presence of spectrum distortions and foreground emissions, the correct way to account for the aberration effect deserves further consideration.
For general purposes, we provided recipes to compute aberration kernels of Doppler weight $d \neq 1$, relevant for boosting, e.g., the specific intensity or the frequency-integrated intensity. Those have been shown to be related to the $d=1$ kernels via $d$-raising/lowering recursions.

Another major result derived from the matrix-element representation is the flow of the $d=1$ aberration kernels with the rapidity parameter $\eta$. This leads to coupled ODEs for a set of aberration kernels that in practice can be effectively truncated. The ODE approach is very advantageous because the initial conditions needed, i.e. the kernels for $\eta=0$, are in all cases trivial, and therefore extremely straightforward to set up. Utilizing standard recipes, the ODE approach can improve upon the recursive approach by a factor $\sim 25$ in terms of computational speed, for moderate values of $\beta \sim 10^{-3}$. Parallelization is straightforward in the ODE approach, pushing the computation of the aberration kernel to a few seconds.

In the limit of large $\ell$, we find simple asymptotic approximations for the kernel elements from the differential equation system (Sec.~\ref{sec:asymp_exp}). While similar to the expressions given earlier by \cite{Notari:2011sb}, we obtain our approximations with purely analytic arguments. Our approximation generally work very well ($\simeq 0.1\%-5\%$ for $\beta =10^{-3}$ and $\ell \leq 4000$), however, when comparing with our ODE approach we find several cases for which the approximation is very far off. 
For $\Delta \ell/\ell \ll 1$, our expressions also capture the main dependence of the kernel even for $\beta\simeq 0.1$; however, since the kernel becomes very wide once $\ell\gg 1/\beta$, the approximation still has limited applicability.
We thus do not recommend using the expression for real computations, also because the ODE approach already is very fast and reliable.

Finally, we emphasize that most of the analytical results obtained in this paper apply to all angular scales $\ell$, arbitrary spin weight $s$ and Doppler weight $d$, being fully non-linear in $\beta$. Therefore, our formalism might find applications in other studies, where anisotropic radiation seen in a (relativistically) boosted reference frame is involved. One example is the scattering of diffuse photon backgrounds by fast-moving charged particles within the jets of active galactic nuclei~\cite{Leismann2005, Mimica2007, Mimica2009}.

\begin{acknowledgments}

The authors thank Donghui Jeong, Marc Kamionkowski and Jared Kaplan for useful discussions. We also thank Miguel Quartin and Alessio Notari for comments on the first draft of the paper. This work was supported by DoE SC-0008108 and NASA NNX12AE86G. 

\end{acknowledgments}

\begin{widetext}

\small

\appendix

\section{Deriving the integral forms for the aberration kernels}
\label{app:integral-form}

In the literature, different approaches of deriving the aberration kernels for both temperature and polarization have been presented (see e.g.~\cite{Challinor:2002zh,Newman:2010gx}). All are essentially based on how the photon's four-momentum and polarization tensor transform under a Lorentz boost of the reference frame. Here we adopt the covariant formalism of Ref.~\cite{Pitrou:2008hy}. 

The photon phase space density needs to be described by a Lorentz tensor $F_{\mu\nu}$ (not to be confused with the usual electromagnetic field-strength tensor), for there are two distinct polarization states. The observer's motion defines a unique time-like unit vector $e^{\mu}_o$, and its line of sight direction (opposite to the direction of propagation) defines a space-like unit vector $n^\mu$ orthogonal to $e^\mu_o$. The symmetric screen-projection tensor can be defined as
\bea
S^\mu{}_\nu(e_o, n) = g^\mu{}_\nu + e^\mu_o e_{o,\nu} - n^\mu n_\nu,
\eea
where $g_{\mu\nu}$ is the flat Minkowski metric. A gauge-invariant phase space density $f_{\mu\nu}$ can be then obtained by screen-projection, i.e. $f_{\mu\nu}=S^\rho{}_\mu S^\sigma{}_\nu F_{\rho\sigma}$. Neglecting circular polarization, which is irrelevant for the CMB, the gauge-invariant $f_{\mu\nu}$ can be decomposed into
\bea
f_{\mu\nu} (E, \vecth{n}) = \frac12 N(E, \vecth{n}) S_{\mu\nu} + P_{\mu\nu}(E, \vecth{n}),
\eea
where $N=g^{\mu\nu} f_{\mu\nu}$ is the occupation number including both polarization states, and the symmetric trace-free $P_{\mu\nu}$ encodes the difference between the two linear polarizations. Note that photon phase space density is a function of the measured photon 4-momentum $p^\mu=E(e^\mu_o - n^\mu)$, or equivalently a function of the measured energy $E$ and the measured line of sight direction $n^\mu$. We have made this dependence manifest.

Next we need to know how $f_{\mu\nu}$ transforms under a Lorentz boost. Note that due to the screen-projection procedure it transforms differently from how a usual Lorentz tensor does. To derive the correct transformation rule, let us consider another observer,
\bea
e^\mu_{o'} = \gamma(e^\mu_o + v^\mu), \qquad \gamma = 1/\sqrt{1-v^\mu v_\mu} = 1/\sqrt{1-\beta^2},\qquad e^\mu_o v_\mu = 0,
\eea 
which has velocity $v^\mu$ relative to the original observer. The new observer will measure screen-projected phase space density $f_{\mu\nu}'=S'^\rho{}_\mu S'^\sigma{}_\nu F_{\rho\sigma}$, where $S'^\mu{}_\nu=S'^\mu{}_\nu(e_{o'},n')$ is the boosted screen-projection tensor constructed from the new time direction $e^\mu_{o'}$ and the aberrated line of sight direction
\bea
n'^{\mu} = - \frac{e^\mu_o - n^\mu}{\gamma(1 + n^\nu v_\nu)} + \gamma (e^\mu_o + v^\mu).
\eea
Besides, the photon energy is shifted to
\bea
\label{eq:energy-shift}
E = E' \gamma (1-n'^\mu v_\mu).
\eea
A nice property is that $S'^\mu{}_\nu$ can be obtained from $S^\mu{}_\nu$ simply through further screen-projections $S'^\mu{}_\nu = S'^\mu{}_\rho S'^\sigma{}_\nu S^\rho{}_\sigma$. After simple algebra this leads to the transformation of $N$ and $P_{\mu\nu}$,
\bea
\label{eq:transformation-intensity-Stokes}
N'(E', \vecth{n}') = N(E, \vecth{n}), \qquad P'_{\mu\nu}(E', \vecth{n}') = S'^\rho{}_\mu S'^\sigma{}_\nu P_{\rho\sigma}(E, \vecth{n}). 
\eea

\subsection{Temperature kernels}

For a blackbody spectrum, we have $N(E,n^\mu)=2/(e^{E/T(\vecth{n})}-1)$. Under a Lorentz boost the blackbody shape is preserved. \refeq{transformation-intensity-Stokes} then implies that $E'/T'(\vecth{n}')=E/T(\vecth{n})$. Using \refeq{energy-shift}, we thus immediately confirm $T'(\vecth{n}')=T(\vecth{n})/[\gamma(1-\beta \cos\theta')]$. We can then compute the boosted temperature multipole coefficients, 
\bea
a^{\prime T}_{l'm'} = \int \id^2 \vecth{n}' Y^*_{l'm'}(\vecth{n}') T'(\vecth{n}') = \int \id^2 \vecth{n}' \frac{Y^*_{l'm'}(\vecth{n}') T(n)}{\gamma \left(1 - n^{\prime\mu}v_{\mu}\right)} = \sum_{lm} a^T_{lm} \int \id^2 \vecth{n}' \frac{Y^*_{l'm'}(\vecth{n}') Y_{lm}(\vecth{n})}{\gamma \left(1 - n^{\prime\mu}v_{\mu}\right)}.
\eea
Given that $n'^\mu v_\mu=\beta\cos\theta'$ and that for boosts along the $z$-direction the azimuthal integral gives $\delta_{mm'}$, together with the definition \refeq{aber-kernel-def} for the aberration kernel, we confirm \refeq{kernel-integral} for $s=0$.

\subsection{Polarization kernels}
\label{app:integral-form-polarization}

We use the notations and properties of tensor spherical harmonics as developed in Ref.~\cite{Dai:2012bc}. We consider the {\it temperature-weighted} polarization tensor
\bea
\mathcal{P}_{\mu\nu}(\vecth{n}) \equiv T(\vecth{n}) P_{\mu\nu}(E, \vecth{n}) / N(E, \vecth{n}).
\eea
Assuming no deviation from a blackbody spectrum, the photon energy $E$ cancels out in $\mathcal{P}_{\mu\nu}$.

Orthogonal to $e^\mu_o$ and $n^\mu$, we can choose two space-like unit vectors $e^\mu_a(\vecth{n})$ with $a=1,2$, which is unaffected by screen projection $e^\mu_a(\vecth{n}) S^\nu{}_\mu(\vecth{n})=e^\nu_a(\vecth{n})$. Since $\mathcal{P}_{\mu\nu}$ is screen-projected, we can construct the two-by-two transverse tensor
\bea
\mathcal{P}_{ab}(\vecth{n}) = e^\mu_a(\vecth{n}) e^\nu_b(\vecth{n}) \mathcal{P}_{\mu\nu}(\vecth{n}).
\eea
We now compute the transformation of $\mathcal{P}_{ab}$
\bea
\mathcal P'_{a'b'}(\vecth{n}') & = & e^{\prime\mu}_{a'}(\vecth{n}') e^{\prime\nu}_{b'}(\hat  n') \mathcal P'_{\mu\nu}(\vecth{n}') = e^{\prime\mu}_{a'}(\vecth{n}') e^{\prime\nu}_{b'}(\vecth{n}') S'_{\mu}{}^{\rho}(\hat  n') S'_{\nu}{}^{\sigma}(\vecth{n}') \frac{T'(\vecth{n}')}{T(\vecth{n})} \mathcal P_{\rho\sigma}(\vecth{n}) \nn\\
& = & \frac{T'(\vecth{n}')}{T(\vecth{n})} e^{\prime\rho}_{a'}(\vecth{n}') e^{\prime\sigma}_{b'}(\vecth{n}') \mathcal P_{\rho\sigma}(\vecth{n}) = \frac{T'(\vecth{n}')}{T(\vecth{n})} e^{\prime\rho}_{a'}(\vecth{n}') e^{\prime\sigma}_{b'}(\vecth{n}') e_{a,\rho}(\vecth{n}) e_{b,\sigma}(\vecth{n}) \mathcal P_{ab}(\vecth{n}).
\eea
The spherical harmonic expansion for the polarization tensor reads
\bea
\mathcal P_{ab}(\vecth{n}) = \sum_{\ell m} \sum_{s=\pm2} a^{s}_{\ell m} Y^{s}_{(\ell m)ab}(\vecth{n}),
\eea
and similar in the boosted frame. Here $Y^{\pm 2}_{(\ell m)ab}(\vecth{n})$ are the tensor spherical harmonics of definite helicity~\cite{Dai:2012bc}. A derivation parallel to that for temperature then gives the tranformation of the multipole coefficients,
\bea
a'^{s'}_{\ell' m'} = \sum_{\ell m} \sum_{s=\pm2} a^s_{\ell' m'} \int \id^2 \vecth{n}' \frac{[Y^{s'}_{(l'm')a'b'}(\vecth{n}')]^* Y^s_{(lm)ab}(\vecth{n})}{\gamma \left(1 - n'^\nu v_\nu \right)} e'^\rho_{a'}(\vecth{n}') e'^\sigma_{b'}(\vecth{n}') e_{a,\rho}(\vecth{n}) e_{b,\sigma}(\vecth{n}).
\eea
We can relate the tensor spherical harmonics to the spin-weighted harmonics $Y^{\pm2}_{(\ell m)ab}(\vecth{n})=\,{}_{\mp2}Y_{(\ell m)}(\vecth{n}) \varepsilon_{\mp2,ab}(\vecth{n})$~\cite{Dai:2012bc}, with $\varepsilon_{\mp2,ab}(\vecth{n})$ being the usual spin-2 base tensors on the sphere. Then we find
\bea
\label{eq:integral-pol-complicated}
a'^{s'}_{\ell' m'} = \sum_{\ell m} \sum_{s=\pm2} a^s_{\ell' m'} \int \id^2 \vecth{n}' \frac{[{}_{-s'}Y_{(\ell' m')}(\vecth{n}')]^* {}_{-s}Y_{(\ell m)}(\vecth{n})}{\gamma \left(1 - n'^\nu v_\nu \right)} [\varepsilon'^{\mu\nu}_{-s'}(\vecth{n}')]^* \varepsilon_{-s,\mu\nu}(\vecth{n}),
\eea
with $\varepsilon^{\mu\nu}_{-s}(\vecth{n})=\varepsilon_{-s,ab}(\vecth{n}) e^\mu_{a}(\vecth{n}) e^\nu_{b}(\vecth{n})$, and similar definition for $\varepsilon'^{\mu\nu}_{-s'}(\vecth{n}')$ as measured in the boosted frame. By explicitly constructing the transverse base vectors $e^\mu_a(\vecth{n})$, and similarly for $e'^\mu_a(\vecth{n}')$ in the boosted frame, one find simple results,
\bea
[\varepsilon'^{\mu\nu}_{\pm2}(\vecth{n}')]^* \varepsilon_{\pm2,\mu\nu}(\vecth{n}) = 1, \qquad [\varepsilon'^{\mu\nu}_{\pm2}(\vecth{n}')]^* \varepsilon_{\mp2,\mu\nu}(\vecth{n}) = 0.
\eea
\refeq{integral-pol-complicated} then simplifies to
\bea
a'^{\pm2}_{\ell' m'} = \sum_{\ell m} a^{\pm2}_{\ell' m'} \int \id^2 \vecth{n}' \frac{[{}_{\mp2}Y_{(\ell' m')}(\vecth{n}')]^* {}_{\mp2}Y_{(\ell m)}(\vecth{n})}{\gamma \left(1 - n'^\nu v_\nu \right)}.
\eea
Again, given that $n'^\mu v_\mu=\beta\cos\theta'$ and that the integral gives $\delta_{mm'}$, together with the definition \refeq{aber-kernel-def} for the aberration kernel, we obtain exactly \refeq{kernel-integral} for $s=\pm2$.

\section{Proof of general kernel properties}
\label{app:proofs}

Here, we briefly prove the general kernel properties, Eq.~\eqref{eq:kernel_props}. We start with $\abkerd{\ell'}{\ell}{-m}{-s}{d}(\beta) = \abkerd{\ell'}{\ell}{m}{s}{d}(\beta)$, which is the simplest to show. From \refeq{kernel-integral_d}, with ${}_{-s}Y_{\ell' -m}(\vecth{n}')=(-1)^{s+m}[{}_{s}Y_{\ell' m}(\vecth{n}')]^*$ we have
\bea
\abkerd{\ell'}{\ell}{-m}{-s}{d}(\beta) = \int \id^2 \vecth{n}' \frac{[{}_{s}Y_{\ell' -m}(\vecth{n}')]^*\,{}_{s}Y_{\ell -m}(\vecth{n})}{[\gamma(1-\beta\cos\theta')]^d}
=  \int \id^2 \vecth{n}' \frac{{}_{-s}Y_{\ell' m}(\vecth{n}')\,[{}_{-s}Y_{\ell m}(\vecth{n})]^*}{[\gamma(1-\beta\cos\theta')]^d}
=[\abkerd{\ell'}{\ell}{m}{s}{d}(\beta)]^\ast\equiv \abkerd{\ell'}{\ell}{m}{s}{d}(\beta),
\eea
which was our statement. Next we prove $\abkerd{\ell'}{\ell}{m}{s}{d}(-\beta) = (-1)^{\ell+\ell'}\abkerd{\ell'}{\ell}{m}{-s}{d}(\beta)$. Imagine we perform a transformation $\theta \rightarrow \pi-\theta$, $\phi\rightarrow \phi+\pi$ and $\beta \rightarrow -\beta$. Then this also means $\theta' \rightarrow \pi-\theta'$ and $\phi'\rightarrow \phi'+\pi$, so that the kernel returns to its initial state. Using the property of spin-weighted harmonics
\bea
{}_s Y_{\ell m}\left(\pi-\theta,\phi+\pi\right) = (-1)^\ell {}_{-s} Y_{\ell m}(\theta,\phi),
\eea
from \refeq{kernel-integral_d} we can directly infer $\abkerd{\ell'}{\ell}{m}{s}{d}(-\beta) = (-1)^{\ell+\ell'}\,\abkerd{\ell'}{\ell}{m}{-s}{d}(\beta)$. This is a symmetry of the kernels for general $d$. 

Finally, we prove $\abkerd{\ell}{\ell'}{m}{s}{d}(\beta) = (-1)^{\ell+\ell'} \abkerd{\ell'}{\ell}{m}{-s}{2-d}(\beta)$, for which we need the identities, $\id \cos \theta'/(1-\beta \cos \theta')=\id \cos \theta/(1+\beta\cos \theta)$, $\id \phi'=\id\phi$ and $1=\gamma^2(1+\beta \cos \theta)(1-\beta\cos \theta')$:
\begin{align}
\abkerd{\ell'}{\ell}{m}{s}{d}(\beta) 
&= \int \id\phi' \,\id \cos \theta' \frac{[{}_{-s}Y_{\ell' m}(\vecth{n}')]^*\,{}_{-s}Y_{\ell m}(\vecth{n})}{[\gamma(1-\beta\cos \theta')]^d}
= \int \id\phi \,\id \cos \theta \frac{[{}_{-s}Y_{\ell' m}(\vecth{n}')]^*\,{}_{-s}Y_{\ell m}(\vecth{n})}{\gamma^d\,(1-\beta\cos \theta')^{d-1}(1+\beta \cos \theta)}
\nn\\
&= \int \id\phi \,\id \cos \theta \frac{\gamma^{2(d-1)}(1+\beta\cos \theta)^{d-1}[{}_{-s}Y_{\ell' m}(\vecth{n}')]^*\,{}_{-s}Y_{\ell m}(\vecth{n})}{\gamma^d(1+\beta \cos \theta)}
=\int \id\phi \,\id \cos \theta \frac{[{}_{-s}Y_{\ell' m}(\vecth{n}')]^*\,{}_{-s}Y_{\ell m}(\vecth{n})}{[\gamma(1+\beta \cos \theta)]^{2-d}}
\nn\\
&=[\abkerd{\ell}{\ell'}{m}{s}{2-d}(-\beta)]^\ast\equiv \abkerd{\ell}{\ell'}{m}{s}{2-d}(-\beta)
\end{align}
from which our statement follows after using $\abkerd{\ell'}{\ell}{m}{s}{d}(-\beta) = (-1)^{\ell+\ell'}\,\abkerd{\ell'}{\ell}{m}{-s}{d}(\beta)$.

\section{Acting $\hat Y_z$ on base states}
\label{app:Yz-action}

To derive \refeq{Yz-action}, we start with the following identities for spin-weighted harmonics~\cite{Varshalovich:book,Challinor:2002zh}
\begin{subequations}
\label{eq:sYlm_relations}
\bea
\label{eq:sYlm_relations_a}
\mu \,{}_sY_{\ell m}(\vecth{n}) 
& = & \sBml{s}{m}{\ell+1}  \,{}_sY_{\ell+1\,m}(\vecth{n}) - \frac{sm}{\ell(\ell+1)}\,{}_sY_{\ell m}(\vecth{n}) 
+ \sBml{s}{m}{\ell} \,{}_sY_{\ell-1\,m}(\vecth{n}), 
\\
\label{eq:sYlm_relations_b}
\sqrt{1-\mu^2} \, \partial_\theta \,{}_sY_{\ell m}(\vecth{n}) 
& = & \ell \sBml{s}{m}{\ell+1} \,{}_sY_{\ell+1\,m}(\vecth{n}) + \frac{sm}{\ell(\ell+1)}\,{}_sY_{\ell m}(\vecth{n}) 
- (\ell+1) \sBml{s}{m}{\ell} \,{}_sY_{\ell-1\,m}(\vecth{n}).
\eea
\end{subequations}
where $\mu=\cos\theta$ and $\sBml{s}{m}{\ell}=\sCml{s}{m}{\ell}/\ell\equiv \sqrt{(\ell^2-m^2)(\ell^2-s^2)/(4\ell^2-1)}/\ell$. From \refeq{Yz-explicit}, we see that $\hat Y_z$ is just the sum of $\cos\theta$ and $\sin\theta\partial_\theta$. When acting on $\ket{s\ell m}$, or explicitly $\tilde D^\ell_{sm}$, we just have to put back the $\psi$-independence $\expf{is\psi}$. The terms proportional to $\ket{s\ell m}$ on the right hand side cancel, and we obtain \refeq{Yz-action}.

\section{Independence of the kernel on the sign of the spin weight $s$ for $d=1$}
\label{app:sign-spinweight}
For convenience, we regard aberration kernels as functions of the rapidity $\eta$. Since $\beta \rightarrow -\beta$ implies $\eta\rightarrow - \eta$, from Eq.~\eqref{eq:kernel_props} we also have $\abkerd{\ell'}{\ell}{m}{s}{d}(-\eta) = (-1)^{\ell+\ell'}\abkerd{\ell'}{\ell}{m}{-s}{d}(\eta)$. By Taylor expanding \refeq{kernel-integral_d} in $\eta$, the kernel for infinitesimal boost reads
\bea
\abkerd{\ell'}{\ell}{m}{s}{d}(\eta) = \delta_{\ell'\ell} + \eta \left[ \frac{\ell+d}{\ell+1} \sCml{s}{m}{\ell+1} \delta_{\ell,\ell'-1} + \frac{(d-1)sm}{\ell(\ell+1)} \delta_{\ell\ell'} - \frac{\ell + 1-d}{\ell} \sCml{s}{m}{\ell} \delta_{\ell,\ell'+1} \right] + \mathcal{O}(\eta^2).
\eea 
The second term in the square brackets, being the only term depending on the sign of $s$ (note that $\sCml{s}{m}{\ell}$ only depends on $|s|$), vanishes if $d=1$. Therefore, we find that at least for infinitesimal boost, \refeq{kernel-symmetry-s} holds if and only if $d=1$. We should expect the $d=1$ case of \refeq{kernel-symmetry-s} is also true for any finite $\eta$, since $\eta$ is additive under successive boosts, and a finite boost is equivalent to many boosts with infinitesimal $\eta$ applied successively. 

To prove that, we write down a Taylor expansion in $\eta$ (specialized to $d=1$),
\bea
\label{eq:taylor-expand-eta-kernel}
\abker{\ell'}{\ell}{m}{s} = \sum^{\infty}_{n=0} \abkappa{\ell'}{\ell}{m}{s}{n} \eta^n.
\eea
We want to show that $\abkappa{\ell'}{\ell}{m}{s}{n}$ is nonzero only if $\ell+\ell'+n$=even. In order to show that, we use $\abker{\ell'}{\ell}{m}{s}=\bra{s \ell' m} \exp(i\eta \hat Y_z) \ket{s \ell m}$, and Taylor-expand the operator
\bea
\exp(i\eta \hat Y_z) = \sum^{\infty}_{n=0} \frac{\eta^n}{n!}  \left(i \hat Y_z \right)^n.
\eea
Then $\abkappa{\ell'}{\ell}{m}{s}{n}=(1/n!)\bra{s\ell' m} ( i \hat Y_z )^n \ket{s\ell m}$. We know that in harmonic space, $\hat Y_z$ has matrix elements only for $\Delta \ell=\ell'-\ell=\pm 1$,
\bea
\bra{s \ell' m} i \hat Y_z \ket{s \ell m} = \sCml{s}{m}{\ell+1} \,\delta_{\ell,\ell'-1} - \sCml{s}{m}{\ell} \,\delta_{\ell,\ell'+1} .
\eea
Then let us examine the matrix elements for $(i \hat Y_z)^n$, which can be obtained by matrix multiplications. Each multiplication changes $\Delta l$ by one unit, therefore $n$ successive multiplications will contribute to a particular value of $\Delta \ell$, only if $\Delta\ell+n=$even, or $\ell+\ell'+n=~$even. Therefore, $\abkappa{\ell'}{\ell}{m}{s}{n}=0$ unless $\ell'+\ell+n=~$even.

Now we can apply the Taylor expansion \refeq{taylor-expand-eta-kernel}, 
\bea
\abker{\ell'}{\ell}{m}{s}(-\eta) = \sum^{\infty}_{n=0} \abkappa{\ell'}{\ell}{m}{s}{n} (-1)^n \eta^n = (-1)^{\ell+\ell'} \sum^{\infty}_{n=0} \abkappa{\ell'}{\ell}{m}{s}{n} \eta^n = (-1)^{\ell+\ell'} \abker{\ell'}{\ell}{m}{s}(\eta).
\eea 
Because $\abkerd{\ell'}{\ell}{m}{s}{d}(-\eta) = (-1)^{\ell+\ell'}\abkerd{\ell'}{\ell}{m}{-s}{d}(\eta)$, which holds for any $d$, we obtain $\abkerd{\ell'}{\ell}{m}{s}{1}(\beta) \equiv \abkerd{\ell'}{\ell}{m}{-s}{1}(\beta)$, as stated in \refeq{kernel-symmetry-s}.  

\section{Numerical computation of $\Kerns{s}{\ell}{m}{\ell'}(\beta)$}
\label{app:sP_lm}
To compute the aberration kernel, $\Kerns{s}{\ell}{m}{\ell'}(\beta)$, we need to evaluate the spin-weighted spherical harmonic functions (here we directly use that the sign of $s$ does not matter), $\sYlm{s}{\ell}{m}(\vecth{n})$. Since we aligned the direction of the motion with the $z$-axis, the dependence on the azimuthal angle, $\phi$, drops out of the problem and for convenience we can introduce the polynomials, 
\begin{align}
\sPlm{s}{\ell}{m}(\cos\theta)=\sqrt{4\pi} \,\expf{- i m \phi} \sYlm{s}{\ell}{m}(\phi, \theta ),
\end{align}
which are real functions. For $s=0$ we have $\sPlm{0}{\ell}{m}(x)=\sqrt{2\ell +1} \sqrt{(\ell-m)!/(\ell+m)!} \,P^m_\ell(x)$, where $P^m_\ell(x)$ define the usual associated Legendre polynomials. With this definition the kernel reads
\begin{align}
\Kerns{s}{\ell}{m}{\ell'}(\beta)=\frac{1}{2}\int_{-1}^1 \frac{\sPlm{s}{\ell'}{m}(\mu')\,\sPlm{s}{\ell}{m}\left(\frac{\mu'-\beta}{1-\beta\mu'}\right)}{\gamma(1-\beta \mu')} \id \mu',
\end{align}
where we used $\mu'=\cos\theta'$. The polynomials $\sPlm{s}{\ell}{m}(x)$ follow the recursion relation
\begin{align}
\sBml{s}{m}{\ell} \sPlm{s}{\ell}{m}(x)=\left(x +\frac{sm}{\ell(\ell-1)}\right)\sPlm{s}{\ell-1}{m}(x)-\sBml{s}{m}{\ell-1} \sPlm{s}{\ell-2}{m}(x),
\end{align}
with $\sBml{s}{m}{\ell}=\sCml{s}{m}{\ell}/\ell\equiv \sqrt{(\ell^2-m^2)(\ell^2-s^2)/(4\ell^2-1)}/\ell$ for $\ell>0$ and $\sBml{s}{m}{\ell}=0$ otherwise. This expression directly follows from Eq.~\eqref{eq:sYlm_relations_a}. The recursions are best started at $\sPlm{s}{m}{m}(x)$. For $s>0$, the initial conditions can be derived by subsequently applying the spin-raising operator, $\eth=-(\sin \theta)^s[\partial_\theta+(i/\sin\theta)\partial_\phi]/(\sin \theta)^s$, to ${}_s Y_{mm}(\phi, \theta)$ and then converting back to $\sPlm{s}{\ell}{m}(x)$. For the first few values of $s$, we find
\begin{subequations}
\label{eq:sPmm_initial}
\begin{align}
\sPlm{0}{m}{m}(x)&=\sqrt{\frac{2m+1}{(2m)!}} P^m_m(x)=(-1)^m  \frac{\sqrt{(2m+1)(2m)!}}{2^m m!} (1-x^2)^{m/2}
\\
\sPlm{1}{1}{0}(x)&=\sqrt{\frac{3}{2}(1-x^2)}, 
\quad 
\sPlm{1}{m}{m}(x)= \sqrt{\frac{m}{m+1}}\sqrt{\frac{1-x}{1+x}}\,\sPlm{0}{m}{m}(x)
\\
\sPlm{2}{2}{0}(x)&=\sqrt{\frac{15}{8}}(1-x^2),
\quad
\sPlm{2}{2}{1}(x)=\sqrt{\frac{5}{4}(1-x^2)}(1-x),
\quad
\sPlm{2}{m}{m}(x)= \sqrt{\frac{m(m-1)}{(m+2)(m+1)}}\frac{1-x}{1+x}\,\sPlm{0}{m}{m}(x).
\end{align}
\end{subequations}
Since $\sPlm{1}{0}{0}(x)=\sPlm{2}{0}{0}(x)=\sPlm{2}{1}{0}(x)=\sPlm{2}{1}{1}(x)=0$, for $s>0$ we need $s$ additional expressions to start the recursions. To carry out the numerical integrals we use Clenshaw-Curtis quadrature rules, which are very accurate even for large $\ell$. We used the results for direct numerical integration to confirm those obtained with the kernel recursion relations.

\end{widetext}




\begin{thebibliography}{99}
\bibliographystyle{unsrt}

\bibitem{Hinshaw:2012aka} 
  G.~Hinshaw {\it et al.}  [WMAP Collaboration],
  Astrophys.\ J.\ Suppl.\  {\bf 208}, 19 (2013)
  [arXiv:1212.5226 [astro-ph.CO]].

\bibitem{Ade:2013zuv} 
  P.~A.~R.~Ade {\it et al.}  [Planck Collaboration],
  arXiv:1303.5076 [astro-ph.CO].

\bibitem[Smoot et al.(1992)]{Smoot1992} 
Smoot, G.~F., Bennett, C.~L., Kogut, A., et al.\ 1992, ApJL, 396, L1 

\bibitem[Mather et al.(1994)]{Mather1994} 
Mather, J.~C., Cheng, E.~S., Cottingham, D.~A., et al.\ 1994, ApJ, 420, 439 

\bibitem{Fixsen:1996nj}
 D.~J.~Fixsen, E.~S.~Cheng, J.~M.~Gales, J.~C.~Mather, R.~A.~Shafer and E.~L.~Wright,
 Astrophys.\ J.\  {\bf 473}, 576 (1996)
 [arXiv:astro-ph/9605054];
 D.~J.~Fixsen,
 Astrophys.\ J.\  {\bf 707}, 916 (2009)
 [arXiv:0911.1955 [astro-ph.CO]].

\bibitem[Fixsen(2009)]{Fixsen2009} 
Fixsen, D.~J.\ 2009, \apj, 707, 916 

\bibitem{dipole} P.~M.~Lubin, G.~L.~Epstein, and G.~F.~Smoot,
     Phys.\ Rev.\ Lett.\ {\bf 50}, 616 (1983); D.~J.~Fixsen,
     E.~S.~Cheng, and D.~T.~Wilkinson, Phys.\ Rev.\ Lett.\
     {\bf 50}, 620 (1983); A.~Kogut {\it et al.}, Astrophys.\ J.\
     {\bf 419} 1 (1993); A.~Kogut {\it et al.},  
     Astrophys.\ J.\  {\bf 419}, 1 (1993)
     [arXiv:astro-ph/9312056].

\bibitem{Hinshaw:2008kr}
 G.~Hinshaw {\it et al.}  [WMAP Collaboration],
 Astrophys.\ J.\ Suppl.\  {\bf 180}, 225 (2009)
 [arXiv:0803.0732 [astro-ph]].

\bibitem{Aghanim:2013suk} 
  N.~Aghanim {\it et al.}  [Planck Collaboration],
  arXiv:1303.5087 [astro-ph.CO].

\bibitem{Challinor:2002zh} 
  A.~Challinor and F.~van Leeuwen,
  Phys.\ Rev.\ D {\bf 65}, 103001 (2002)
  [astro-ph/0112457].

\bibitem{Kosowsky:2010jm} 
  A.~Kosowsky and T.~Kahniashvili,
  Phys.\ Rev.\ Lett.\  {\bf 106}, 191301 (2011)
  [arXiv:1007.4539 [astro-ph.CO]].

\bibitem{Amendola:2010ty} 
  L.~Amendola, R.~Catena, I.~Masina, A.~Notari, M.~Quartin and C.~Quercellini,
  JCAP {\bf 1107}, 027 (2011)
  [arXiv:1008.1183 [astro-ph.CO]].

\bibitem{Burles:2006xf} 
  S.~Burles and S.~Rappaport,
  Astrophys.\ J.\ Lett.\ {bf 641}, L1 (2006)
  [arXiv:astro-ph/0601559].

\bibitem{Jeong:2013sxy} 
  D.~Jeong, J.~Chluba, L.~Dai, M.~Kamionkowski and X.~Wang,
  Phys.\ Rev.\ D {\bf 89}, 023003 (2014)
  [arXiv:1309.2285 [astro-ph.CO]].

\bibitem{Catena:2012hq} 
  R.~Catena and A.~Notari,
  JCAP {\bf 1304}, 028 (2013)
  [arXiv:1210.2731 [astro-ph.CO]].

\bibitem{Pereira:2010dn} 
  T.~S.~Pereira, A.~Yoho, M.~Stuke and G.~D.~Starkman,
  arXiv:1009.4937 [astro-ph.CO].

\bibitem{Jeong:future}
  D.~Jeong {\it et al.}, in preparation.

\bibitem{Ade:2013kta} 
  P.~A.~R.~Ade {\it et al.}  [Planck Collaboration],
  arXiv:1303.5075 [astro-ph.CO].

\bibitem{Story:2012wx} 
  K.~T.~Story, C.~L.~Reichardt, Z.~Hou, R.~Keisler, K.~A.~Aird, B.~A.~Benson, L.~E.~Bleem and J.~E.~Carlstrom {\it et al.},
  Astrophys.\ J.\  {\bf 779}, 86 (2013)
  [arXiv:1210.7231 [astro-ph.CO]].

\bibitem{Fowler:2010cy} 
  J.~W.~Fowler {\it et al.}  [ACT Collaboration],
  Astrophys.\ J.\  {\bf 722}, 1148 (2010)
  [arXiv:1001.2934 [astro-ph.CO]].

\bibitem{Niemack:2010wz} 
  M.~D.~Niemack, P.~A.~R.~Ade, J.~Aguirre, F.~Barrientos, J.~A.~Beall, J.~R.~Bond, J.~Britton and H.~M.~Cho {\it et al.},
  Proc.\ SPIE Int.\ Soc.\ Opt.\ Eng.\  {\bf 7741}, 77411S (2010)
  [arXiv:1006.5049 [astro-ph.IM]].

\bibitem{Austermann:2012ga} 
  J.~E.~Austermann, K.~A.~Aird, J.~A.~Beall, D.~Becker, A.~Bender, B.~A.~Benson, L.~E.~Bleem and J.~Britton {\it et al.},
  Proc.\ SPIE Int.\ Soc.\ Opt.\ Eng.\  {\bf 8452}, 84521E (2012)
  [arXiv:1210.4970 [astro-ph.IM]].

\bibitem{Notari:2011sb}
 A.~Notari and M.~Quartin,
 JCAP {\bf 1202}, 026 (2012)
 [arXiv:1112.1400 [astro-ph.CO]].

\bibitem{Yoho:2012am}
 A.~Yoho, C.~J.~Copi, G.~D.~Starkman and T.~S.~Pereira,
 Mon.\ Not.\ R.\ Astron.\ Soc.\ {\bf 432}, 2208 (2013)
 [arXiv:1211.6756 [astro-ph.CO]].

\bibitem[Menzies \& Mathews(2005)]{Menzies2005} 
Menzies, D., \& Mathews, G.~J.\ 2005, ApJ, 624, 7 

\bibitem{Chluba:2011zh} 
  J.~Chluba,
  MNRAS, 415, 3227 (2011)
  arXiv:1102.3415 [astro-ph.CO].

\bibitem{Goldberg:1967spin-s}
  J.~N.~Goldberg, A.~J.~Macfarlane, E.~T.~Newman, F.~Rohrlich and E.~C.~G.~Sudarshan,
  J.\ Math.\ Phys. {\bf 8}, 2155 (1967). 

\bibitem{Dai:2012bc} 
  L.~Dai, M.~Kamionkowski and D.~Jeong,
  Phys.\ Rev.\ D {\bf 86}, 125013 (2012)
  [arXiv:1209.0761 [astro-ph.CO]].

\bibitem[Lewis \& Challinor(2006)]{Lewis2006} 
Lewis, A., \& Challinor, A.\ 2006, Physic Reports, 429, 1 

\bibitem[Challinor \& Lewis(2005)]{Challinor2005} 
A.~Challinor, \& A.~Lewis, \prd, 71, 103010 (2005) 

\bibitem[Mimica et al.(2007)]{Mimica2007} 
Mimica, P., Aloy, M.~A., {Mue}ller, E.\ 2007, A\&A, 466, 93 

\bibitem[Leismann et al.(2005)]{Leismann2005} 
T.~Leismann, L.~Anton, M.~A.~Aloy, E.~Muller, J.~M.~Marti, J.~A.~Miralles, J.~M.~Ibanez,\ 2005, A\&A, 436, 503 

\bibitem[Mimica et al.(2009)]{Mimica2009} 
P.~Mimica, M.~A.~Aloy, I.~Agudo, J.~M.~Marti, J.~L.~Gomez, J.~A.~Miralles,\ 2009, \apj, 696, 1142 

\bibitem{Newman:2010gx} 
  E.~T.~Newman and R.~H.~Price,
  Phys.\ Rev.\ D {\bf 82}, 083516 (2010)
  [arXiv:1007.4351 [gr-qc]].

\bibitem{Pitrou:2008hy} 
  C.~Pitrou,
  Class.\ Quant.\ Grav.\  {\bf 26}, 065006 (2009)
  [arXiv:0809.3036 [gr-qc]].

\bibitem{Varshalovich:book}
   D.~A.~Varshalovich, A.~N.~Moskalev, and V.~K.~Khersonskii,
   {\it Quantum Theory of Angular Momentum}
   (World Scientific, Singapore, 1988).








\end{thebibliography}
\end{document}